\documentclass[11pt]{article}

\usepackage[margin=1in]{geometry}
\usepackage[utf8]{inputenc}
\usepackage[T1]{fontenc}
\usepackage{graphicx}
\usepackage{amsmath}
\usepackage{amssymb}
\usepackage{bm}
\usepackage{booktabs}
\usepackage{array}
\usepackage{authblk}
\usepackage{hyperref}

\newenvironment{acknowledgments}{\section*{Acknowledgments}}{}

\title{A minimal dynamical model of human-AI cognitive coevolution}

\author[1,*]{Petar Jolakoski}
\author[1,2,3]{Dimitar Trajanov}
\author[4]{Qiyu Sun}
\author[5]{Zhen Wu}
\author[1,2]{Ljupco Kocarev}

\affil[1]{Macedonian Academy of Sciences and Arts, Skopje, Macedonia}
\affil[2]{Faculty of Computer Science and Engineering, Ss. Cyril and Methodius University, Skopje, Macedonia}
\affil[3]{Metropolitan College, Boston University, Boston, USA}
\affil[4]{School of Information Science and Engineering, East China University of Science and Technology, Shanghai 200237, China}
\affil[5]{Department of Psychological and Cognitive Sciences, Tsinghua University, Beijing 100084, China}
\affil[*]{Corresponding author: \href{mailto:jolakoskip@manu.edu.mk}{jolakoskip@manu.edu.mk}}

\date{September 20, 2026}

\begin{document}

\maketitle

\begin{abstract}

The rapid deployment of generative AI raises the question of whether pervasive cognitive offloading drives a measurable, potentially irreversible decline in unassisted human reasoning.
We examine this question with a phenomenological dynamical model based on the Cattell-Horn-Carroll distinction between fluid intelligence ($G_f$) and crystallized intelligence ($G_c$).
Starting from the discrete-time Bivariate Latent Change Score formalism used in cognitive-aging research, we derive its continuous-time linear counterpart and introduce a fast third variable representing composite machine-intelligence capability. The machine variable enters the human subsystem through Hill-function modifications of the self-feedback and cross-coupling coefficients.
We emphasize that our phenomenological model captures both the positive and negative effects of AI on human cognition. By formalizing the qualitative concepts of positive and negative impacts, we introduce definitions for AI-induced cognitive decline and AI-induced cognitive augmentation.
A singular-perturbation reduction, justified via Tikhonov's theorem, collapses the resulting three-dimensional system onto a two-dimensional slow manifold. On this manifold we locate two bifurcations of the equilibrium.
In one parameter regime, weakening the contribution of fluid reasoning to crystallized knowledge produces a saddle-node bifurcation that eliminates the stable positive equilibrium.
In another, reversing the contribution of crystallized knowledge to fluid reasoning creates opposing feedback; increasing fluid self-feedback then reduces damping and produces a supercritical Hopf bifurcation with a stable periodic orbit.
We discuss limitations and future directions, interpret our results as a phenomenological account of a coupled socio-technical system, and state the conditions under which each dynamical regime applies.

\end{abstract}

\section{Introduction}
\label{sec:introduction}

Generative artificial intelligence (AI) increasingly allows users to delegate analytical, linguistic, and computational tasks that previously required sustained cognitive effort. Existing studies primarily measure immediate performance and short-term retention, leaving the long-run consequences of repeated offloading unresolved. This gap motivates a coupled, nonlinear representation of human and machine capability, whose feedback we examine as a modeling hypothesis.

Central to this coevolutionary process is cognitive offloading. In human-computer interaction, delegating routine tasks to external devices can improve short-term productivity. Cognitive epidemiology and neuroscience, however, suggest that maintaining cognitive reserve and structural brain plasticity depends heavily on desirable difficulties, that is, the intrinsic mental friction and metabolic effort required during active problem-solving. Pervasive automation may reduce opportunities for effortful practice, with consequences for unassisted task performance and skill retention. 
Early AI studies report cross-sectional associations and short-term, task-specific differences in reasoning, immediate recall, or engagement, but they do not establish persistent decline after assistance is removed; the multi-decade consequences of sustained cognitive offloading remain an open question~\cite{gerlich2025aitools,kosmyna2025brainchatgptaccumulationcognitive,znidaric2026review}.

Evaluating these long-term dynamics requires a formal mathematical framework capable of compressing multi-scale neural and behavioral phenomena into macroscopic state variables. Traditional paradigms in cognitive psychology, such as cognitive load theory or working memory models, primarily address task-scale resource allocation and transient performance, whereas a bifurcation analysis requires long-horizon, mutually coupled latent-state dynamics instead. 
Quantitative psychology offers a tractable state-space representation for this purpose through the Theory of Fluid and Crystallized Intelligence, which distinguishes reasoning in novel situations with comparatively limited reliance on acquired knowledge ($G_f$) from knowledge and skills developed through learning and experience ($G_c$)~\cite{cattell1963theory}.
For tractability, the model projects cognition onto these two broad latent dimensions; the wider Cattell-Horn-Carroll (CHC) taxonomy and its neural implementation remain outside the state space. Historically, these two dimensions have followed distinct, decoupled lifespan trajectories: fluid capacities peak in early adulthood and decline continuously thereafter, while crystallized knowledge remains stable or continues to expand well into senescence.

These differences in lifespan trajectories provide a basis for modeling human-AI cognitive coevolution: it allows technological perturbations to be represented through changes in one or both dimensions, and the existing apparatus for modeling their joint dynamics, the Bivariate Latent Change Score framework, supplies a longitudinal, discrete-time template that can be embedded in continuous time under time-homogeneous assumptions (Sec.~\ref{subsec:fluid_crystallized_theory}). What that framework does not supply is a mechanism by which an external, rapidly evolving technological variable feeds back into the coefficients governing human cognitive change, nor an account of how the resulting coupled system behaves away from its benchmark-based reference point. This paper examines one phenomenological closure of that system. We adapt the fluid-crystallized formalism into a three-variable dynamical system in which a fast machine-intelligence variable modulates the self-feedback and cross-coupling terms of the slow human subsystem (Sec.~\ref{sec:model_formulation}). The resulting model captures both the positive and negative effects of AI on human cognition, and we use it to define AI-induced cognitive decline and AI-induced cognitive augmentation (Sec.~\ref{sec:augmentation-outcomes}). Reducing the singularly perturbed system via Tikhonov's theorem, we then use bifurcation analysis to ask under what conditions cognitive offloading produces smooth decline, node-saddle coexistence, a saddle-node loss, or a Hopf bifurcation of the reference equilibrium  (Sec.~\ref{sec:results}). Section~\ref{sec:disc-and-conc} concludes the paper: it discusses limitations and future directions, interprets the results as a phenomenological account of a coupled socio-technical system, and states the conditions under which each dynamical regime applies.

\section{Methodological Framework and Model Selection}
\label{sec:preliminaries}

This section establishes the theoretical foundation for modeling human-AI cognitive coevolution. More specifically, we explain why fluid and crystallized intelligence provide a suitable level of abstraction for modeling long-term human-AI cognitive dynamics. We review evidence for biological and AI-induced cognitive decline, evaluate current cognitive models, and detail how micro-scale dynamics shape human cognition. To formalize multi-scale neural dynamics without microscopic complexity, we leverage the theory of fluid and crystallized intelligence as a tractable, low-dimensional state-space formalism. By mapping the divergent trajectories of abstract reasoning and accumulated knowledge via coupled differential equations, this approach provides a formal mathematical framework to analyze the long-term dynamics of human-AI interactions.

\subsection{From Biological to AI-Induced Cognitive Decline}
\label{subsec:biological_vs_ai_decline}

Biomedical research on cognitive decline tracks age-related changes in working memory, executive function, and attentional control. A central goal in this field is distinguishing normal aging from clinical diseases like Alzheimer's disease~\cite{petersen1999mild, sperling2011toward}. Researchers map these changes using neuroimaging biomarkers and cerebrospinal-fluid measures to detect brain alterations before symptoms appear~\cite{jack2010hypothetical}. 
Epidemiological syntheses identify potentially modifiable factors associated with dementia risk across the life course, including limited education, hypertension, obesity, diabetes, physical inactivity, smoking, hearing loss, social isolation, and other exposures. Their estimated population-attributable fractions are conditional on causal and modifiability assumptions and should not be read as observed intervention effects~\cite{livingston2020dementia,livingston2024dementia}. Prevention research evaluates individual and policy interventions intended to reduce risk or delay impairment. Longitudinal observational studies also illustrate heterogeneity in later-life outcomes: Levy and Slade found that participants who initially held more positive views of aging had higher odds of later cognitive improvement, whereas Vidal-Pi\~neiro et al. found that greater-than-peer structural brain decline covaried with episodic-memory decline, although the explained variance was small~\cite{levy2026aging,vidal2025vulnerability}.

\v{Z}nidari\v{c} et al. review evidence suggesting that generative AI may reduce effort during cognitively intensive activities such as writing, coding, and reasoning. They emphasize that AI-specific evidence remains dominated by short-term and correlational studies~\cite{znidaric2026review}.
Gerlich found cross-sectional associations between heavier AI use, greater offloading, and lower self-reported or measured critical-thinking outcomes~\cite{gerlich2025aitools}. Kosmyna et al. found that LLM-assisted essay writing caused poorer immediate quote recall and lower EEG task connectivity than searching or writing unaided~\cite{kosmyna2025brainchatgptaccumulationcognitive}.
These findings motivate the study of offloading and deskilling, but they do not establish persistent neuroplastic change. Appendix~\ref{sec:appendix_evidence_offloading} summarizes recent surveys and behavioral experiments documenting task offloading and deskilling in professional and student groups. Appendix~\ref{sec:appendix_atrophy_evidence} outlines the associated neurophysiological markers, metacognitive inflation metrics, and network connectivity changes that have been proposed as possible indicators of such change.

Cognitive epidemiology reports associations between early-life intelligence, lifespan, and long-term health~\cite{calvin2011intelligence, martin2004cognitive}. If automated offloading were shown to alter cognitive development or brain aging at scale, it could therefore have population-health implications; current evidence does not establish that pathway. 
Neuroimaging reviews report structural and functional brain differences associated with Internet Gaming Disorder, especially in executive-control, reward, decision-making, and sensorimotor systems~\cite{weinstein2017new,weinstein2020neurobiological}. These findings concern problematic gaming and do not establish that ordinary smart-device use, multitasking, or technological offloading causes comparable brain changes. Evidence concerning the specific long-term cognitive effects of generative AI remains limited, with longitudinal physiological studies especially scarce~\cite{znidaric2026review}.

Together, this evidence defines the empirical problem and motivates the comparison of cognitive frameworks suitable for modeling its long-term dynamics.

\subsection{Cognitive Models and Models of Cognitive Decline}
\label{subsec:cognitive_models}

Modern cognitive psychology relies on several core theoretical frameworks to quantify information processing, learning constraints, and decision-making. Cognitive Load Theory proposes that learning can be impaired when task demands exceed the limited capacity of working memory, particularly when instructional design imposes avoidable extraneous processing demands~\cite{sweller1988cognitive,sweller2011cognitive}. Baddeley and Hitch’s model conceptualizes working memory as a multicomponent system containing specialized verbal and visuospatial components coordinated by a central executive~\cite{baddeley1974working}. Building on limited-capacity and dual-channel assumptions, the Cognitive Theory of Multimedia Learning proposes that learners can understand material more effectively from appropriately integrated words and pictures than from words alone, although the benefit depends on how the material is designed~\cite{mayer2001multimedia}.

Dual-process theories distinguish relatively rapid, autonomous Type 1 processing from more deliberative Type 2 processing that depends heavily on working memory. These categories describe families of processes rather than two completely separate cognitive systems~\cite{kahneman2011thinking, evans2013dual}. For two-choice decisions, the Drift-Diffusion Model represents responding as noisy evidence accumulation toward one of two decision boundaries~\cite{ratcliff1978theory}.

The Theory of Fluid and Crystallized Intelligence distinguishes the capacity to reason in novel situations from knowledge and skills acquired through learning and experience~\cite{cattell1963theory}.
Theories of cognitive aging address different mechanisms: Salthouse emphasizes processing-speed limitations, Hasher and Zacks emphasize reduced inhibitory efficiency and competition in working memory, and STAC-r describes compensatory neural scaffolding together with life-course enrichment and depletion~\cite{salthouse1996processing,hasher1988working,reuter2014how}.

To relate these cognitive frameworks to the biological mechanisms that may support long-term cognitive change, we next consider the functional and structural dimensions of neuroplasticity.

\subsection{Neuroplastic Mechanisms: Functional and Structural Dimensions} \label{subsec:neuroplastic_mechanisms}

Neuroplasticity is the ability of the central nervous system to reorganize itself in response to internal or external demands~\cite{pascual2005plastic}. 
Neuroplasticity includes activity-dependent changes in synaptic efficacy as well as changes in neural structure. Synaptic plasticity comprises multiple circuit- and context-dependent mechanisms involving transmitter release, receptor trafficking, intracellular signaling, and other presynaptic and postsynaptic processes~\cite{citri2008synaptic,malenka2004ltp}.
In cultured hippocampal neurons, Bi and Poo found that the direction and magnitude of persistent synaptic modification depended on the relative timing of presynaptic and postsynaptic activity~\cite{bi1998synaptic}; this biological experiment does not itself specify a computational-network weight-update rule.
Song et al. supplied a computational proof of principle by implementing pairwise spike-timing-dependent plasticity as bounded excitatory-conductance updates in an integrate-and-fire neuron. Under a slightly depression-dominated learning window, the simulated synapses competed for control of postsynaptic spike timing and approached a balanced, irregular-firing regime~\cite{song2000competitive}. Structural plasticity, by contrast, involves physical changes to neural architecture, including the generation of new cellular components and the reconfiguration of macroscopic networks~\cite{holtmaat2009experience}. At the micro-scale, this process involves dendritic spine remodeling, synaptogenesis, and axonal sprouting. In computational structural-plasticity models, synapse formation and deletion can be represented as changes in network connectivity. Butz et al. showed in a 400-neuron simulation that a homeostatic activity rule combined with distance-dependent connection formation can reorganize network topology~\cite{butz2014homeostatic}. Graph-theoretic measures can then characterize segregation, integration, centrality, and related properties of structural or functional connectivity networks~\cite{rubinov2010complex}.

Longer-term mental training can be associated with changes in MRI-derived cortical morphology, although such changes do not identify the underlying cellular mechanism and do not by themselves establish a sequence from persistent functional activity to molecular cascades and structural remodeling~\cite{valk2017structural}. At the macro scale, anatomical connectivity constrains modeled neural dynamics, while functional connectivity describes statistical dependence among activity signals~\cite{park2013structural,deco2014how,honey2009predicting}. These measurements can inform structure--dynamics models but do not by themselves identify synaptic remodeling or a learning mechanism.

These micro-scale dynamics shape macroscopic organization and human cognition. Structural connectivity (SC) and functional connectivity (FC) provide complementary macro-scale descriptions of anatomical organization and statistical dependence between regional activity signals, respectively~\cite{sporns2005human,vandenheuvel2010exploring}. They are not direct macro-scale measures of structural and functional plasticity. In 415 healthy adults aged 22--37, resting-state FC models had higher median explained variance than SC models across two parcellations and crystallized, fluid, and total cognition, although direct model-comparison tests did not establish significant FC superiority and the models were not externally validated~\cite{dhamala2021distinct}.

The FC and SC models also yielded different feature-importance patterns associated with cognition, whereas feature patterns were strongly correlated across cognitive outcomes within each modality~\cite{dhamala2021distinct}. Computational whole-brain models can instead use anatomical connectivity to constrain neural dynamics and examine how resting-state functional connectivity emerges from the interaction of structure, dynamics, and noise~\cite{deco2013resting}.

\subsection{The Theory of Fluid and Crystallized Intelligence: State-Space Formalization}
\label{subsec:fluid_crystallized_theory}

To represent these multiscale cognitive and neural processes at a tractable level, we use the fluid--crystallized distinction introduced in Sec.~\ref{sec:introduction} as a two-dimensional state-space representation.
Within this representation, fluid intelligence ($G_f$) denotes reasoning in novel situations with limited reliance on acquired knowledge, while crystallized intelligence ($G_c$) denotes knowledge and skills developed through learning and experience~\cite{cattell1963theory}.
The modeled contribution of $G_f$ to $G_c$ follows Cattell's Investment Hypothesis, which proposes that fluid ability contributes to the learning processes through which crystallized knowledge develops~\cite{cattell1963theory}.

In contemporary CHC theory, fluid reasoning, short-term or working memory, processing speed, long-term storage and retrieval, and sensory processing are related but distinguishable broad abilities~\cite{mcgrew2009chc}. For the dynamical model, $G_f$ and $G_c$ serve as broad latent coordinates; the resulting two-axis state space is a dimensional reduction, not the full CHC taxonomy.

Longitudinal structural equation models, including latent-growth and change models, are used to estimate changes in cognitive abilities across age ~\cite{mcardle2000modeling,mcardle2002comparative}. Longitudinal models can estimate within-person change, but they do not automatically separate aging, age-cohort differences, and retest gains. In simulations with true retest effects fixed at zero, omitting cohort effects produced apparent retest effects, while flexible occasion-specific retest terms sharply reduced power to detect age slopes; Hoffman et al. recommend designs that identify short-term retest separately from long-term change~\cite{hoffman2011confounds}. Estimates also remain sensitive to attrition, measurement non-invariance, and model specification. Healthy cognitive-aging trajectories are heterogeneous, and chronological age alone is a poor predictor of individual cognitive outcome~\cite{abrous2026hallmarks}. Lifespan studies indicate different average trajectories for these abilities: fluid reasoning generally peaks earlier and declines earlier, whereas crystallized knowledge peaks later and remains comparatively stable for longer, although both may decline in later life~\cite{mcardle2000modeling, mcardle2002comparative, salthouse2004what}.

McArdle et al. ~\cite{mcardle2000modeling} used longitudinal structural equation models and linear dynamic models to examine growth, decline, and lagged relationships among intellectual
abilities. Latent Change Score models express change through discrete-time difference equations; under appropriate time-homogeneity and parameter-rescaling assumptions, corresponding continuous-time representations can be written as differential equations~\cite{cancer2021dynamical}. For example, Handing et al.~\cite{handing2023age} estimated a substantially steeper average decline in general fluid cognition from ages 70-85 than from ages 50-70, illustrating that a single linear rate need not describe the entire observed age range. A dynamic differential equation framework is a useful candidate for examining conditional trajectories under hypothesized sociotechnical perturbations.

\subsection{Establishing a Framework for Modeling Human-AI Cognitive Coevolution} \label{sec:establishing}

CHC theory provides a broad working taxonomy of human cognitive abilities and has influenced psychological and educational assessment~\cite{mcgrew2009chc}. Hendrycks et al. adapt CHC-inspired domains into a proposed ten-domain framework for evaluating machine capability~\cite{hendrycks2025definitionagi}. We further reduce those domains to two cognitive dimensions. The two-dimensional representation described in Sec.~\ref{subsec:fluid_crystallized_theory} provides a tractable basis for modeling human-AI cognitive coevolution.

Long-term longitudinal evidence on persistent cognitive or neural changes associated with generative-AI use remains limited~\cite{znidaric2026review}.
Kosmyna et al. followed 54 adults through three 20-minute essay-writing sessions conducted over four months. Participants in the LLM-assisted condition showed poorer immediate quotation recall and lower task-related EEG connectivity than the comparison groups; 18 volunteers subsequently completed a condition-switch session~\cite{kosmyna2025brainchatgptaccumulationcognitive}. These findings concern task-related behavior and neural engagement and do not establish permanent structural neural degradation.
Thus, irreversible AI-induced cognitive decline remains a modeling hypothesis and an open empirical question.

Longitudinal aging studies provide evidence of within-person structural brain change. Internet Gaming Disorder (IGD) evidence, by contrast, provides an associative comparison rather than a firm causal baseline for technological disuse. O’Sullivan et al. compared 20 older adults with 10 younger controls using diffusion-tensor MRI. Older participants showed reduced diffusion anisotropy and increased mean diffusivity, with the largest differences in anterior white matter. Within the older group, anterior mean diffusivity was correlated with executive-function performance~\cite{osullivan2001evidence}.

Complementing these connectivity shifts, Raz et al. tracked structural brain changes in healthy adults over a five-year period~\cite{raz2005regional}. They used latent difference score modeling to measure both the average rate of volumetric shrinkage across multiple brain regions and individual differences in those trajectories~\cite{raz2005regional}. Separately, narrative reviews of IGD report heterogeneous structural and functional differences in executive-control, reward, decision-making, emotional, and sensorimotor systems~\cite{weinstein2017new,weinstein2020neurobiological}. Because much of this evidence is cross-sectional and the direction of white-matter findings is not uniform, it does not establish that technological immersion physically alters or degrades neural pathways through the disuse mechanism represented by the model.

The limited evidence on long-term human-AI cognitive change motivates the phenomenological model developed below, which examines possible feedback structures under explicit assumptions.

\section{Model Formulation} \label{sec:model_formulation}

Building on the two-dimensional representation established in Sec.~\ref{subsec:fluid_crystallized_theory}, we denote fluid intelligence by $X$ and crystallized intelligence by $Y$.

Empirical studies describe distinct lifespan trajectories across cognitive abilities. Duncan et al.~\cite{duncan2017complexity} provide evidence that performance on fluid-intelligence matrix problems depends partly on cognitive segmentation, in which complex problems are decomposed into simpler, separately attended parts managed by frontal and parietal control systems. Hartshorne and Germine found that different cognitive abilities, including processing speed, working memory, and abstract reasoning, reach peak performance at different ages~\cite{hartshorne2015when}. CHC theory treats these as related but distinguishable abilities, so their trajectories do not make processing speed and working memory constituent sub-capacities of $G_f$~\cite{mcgrew2009chc}.

Specifically, raw information-processing speed peaks around ages 18--19 and declines immediately thereafter. Short-term and working memory reach maximum capacity between ages 25 and 29. Visual-spatial working memory and complex abstract reasoning plateau and peak later, around ages 30--40~\cite{hartshorne2015when}. By comparison, $G_{c}$ accumulates or remains stable throughout adulthood and typically peaks between ages 60 and 70 before exhibiting minimal late-life decline. $G_{f}$, in contrast, exhibits a continuous decline that accelerates in older age~\cite{murman2015impact}.

\subsection{Bivariate Latent Change Score Formulation} \label{sub-sec:biv}

The methodology of utilizing coupled dynamic systems to model the temporal behavior of Fluid and Crystallized Intelligence is foundational in quantitative psychology and cognitive aging. The primary scientific frameworks formalizing the relationship between $G_{f}$  and $G_{c}$  rely on discrete-time Latent Change Score (LCS) models, pioneered by McArdle et al.~\cite{mcardle2000modeling, mcardle2002comparative}, as discussed in the previous section.

For the Bivariate Latent Change Score (BLCS) model~\cite{cancer2021dynamical}, the equations describing the dynamic changes over time for variable $X$ (fluid intelligence) and variable $Y$ (crystallized intelligence) are given by:
\begin{align}
    \Delta X_{t} &= \alpha_{x} \cdot s_{x} + \beta_{x,\text{disc}} \cdot X_{t-1} + \gamma_{xy,\text{disc}} \cdot Y_{t-1} \label{eq:disc-x} \\
    \Delta Y_{t} &= \alpha_{y} \cdot s_{y} + \beta_{y,\text{disc}} \cdot Y_{t-1} + \gamma_{yx,\text{disc}} \cdot X_{t-1} \label{eq:disc-y} 
\end{align}
Here latent states $X_{t-1}$ and $Y_{t-1}$ are model-estimated levels of fluid intelligence ($X$) and crystallized intelligence ($Y$) at the preceding iteration, conditional on the specified measurement model. Latent change scores $\Delta X_{t}$ and $\Delta Y_{t}$ represent model-implied within-person change in each domain between iterations $t-1$ and $t$. By definition in Latent Change Score modeling, the current state is the sum of the prior state and the change: $X_t = X_{t-1} + \Delta X_t$. 

Constant change factors $s_x$ and $s_y$, scaled by $\alpha_x$ and $\alpha_y$, provide additive components at each change occasion. Their long-run meaning depends jointly on the self-feedback, cross-coupling, and implied equilibrium rather than representing an independent linear rate in isolation. Slope loading coefficients $\alpha_{x}$ and $\alpha_{y}$ are scaling constants that map the constant latent change factors to the respective change scores. In standard applications, these are typically fixed to $1.0$.

Self-feedback (proportional-change) parameters $\beta_{x,\text{disc}}$ and $\beta_{y,\text{disc}}$ represent time-ordered predictive associations between a variable's prior level and its subsequent latent change~\cite{cancer2021dynamical}. A negative value contributes a regressive term, but growth, decline, and the equilibrium cannot be inferred from its sign without also considering the additive and coupling terms. The illustrative calibration specifies different self-feedback values and signs for $X$ and $Y$.

Coupling (cross-lagged) coefficients $\gamma_{xy,\text{disc}}$ and $\gamma_{yx,\text{disc}}$ represent time-ordered predictive associations between the two cognitive domains; BLCS coupling is not by itself a causal effect~\cite{cancer2021dynamical}.
$\gamma_{xy,\text{disc}}$ captures the proportional effect of prior crystallized intelligence ($Y_{t-1}$) on the structural change in fluid intelligence ($\Delta X_t$). $\gamma_{yx,\text{disc}}$ captures the proportional effect of prior fluid intelligence ($X_{t-1}$) on the structural change in crystallized intelligence ($\Delta Y_t$). 
The coupling parameters encode how the prior level of one domain predicts subsequent modeled change in the other. For this phenomenological model, we interpret $\gamma_{yx,\text{disc}}$ as an investment relation and $\gamma_{xy,\text{disc}}$ as a scaffolding relation, and specify their positive signs and relative magnitudes through the calibration.

For the adult-aging illustration, we set the following parameter intervals to permit negative self-feedback, positive cross-coupling, stable equilibria, relaxation times of 20--60 annual steps, and feasible presentation-scale scores. These are calibration ranges, not confidence intervals:
\begin{alignat}{2}
\beta_{x,\text{disc}}   &\in (-0.3,\,0),   &\quad \beta_{y,\text{disc}}   &\in (-0.15,\,0), \label{eq:par-beta}\\
\gamma_{xy,\text{disc}} &\in (0,\,0.15),   &\quad \gamma_{yx,\text{disc}} &\in (0,\,0.3),   \label{eq:par-gamma} \\
\beta_{x,\text{disc}}   &< \beta_{y,\text{disc}}, &\quad \gamma_{xy,\text{disc}}  &< \gamma_{yx,\text{disc}}. \label{eq:par-mix}
\end{alignat}

\subsection{Continuous-Time Formulation of the Bivariate Dynamics} \label{subsect-cont}

For a time-homogeneous linear system, interval-specific discrete transition matrices relate to an underlying continuous-time drift matrix through the matrix exponential, and the discrete coefficients depend on the observation interval~\cite{voelkle2012sem}. We apply this relationship to the deterministic drift component and define $\beta_{\cdot,\text{disc}}$ and $\gamma_{\cdot,\text{disc}}$ over the interval $\Delta t$. 
Let $X(t)$ represent fluid intelligence and $Y(t)$ represent crystallized intelligence as continuous functions of time $t$. The instantaneous rates of change are expressed as the following system of ODEs:
\begin{align}
    \frac{dX(t)}{dt} = \mu_x  + \beta_{x,\text{cont}} \cdot X(t) + \gamma_{xy,\text{cont}} \cdot Y(t) \label{eq:cont-x} \\
    \frac{dY(t)}{dt} = \mu_y  + \beta_{y,\text{cont}} \cdot Y(t) + \gamma_{yx,\text{cont}} \cdot X(t) \label{eq:cont-y}
\end{align}
Continuous self-feedback parameters $\beta_{x,\text{cont}}$ and $\beta_{y,\text{cont}}$ are diagonal entries of the continuous drift matrix. A negative entry contributes local damping, but its substantive interpretation depends on the full coupled system and its constant inputs.
Continuous coupling parameters $\gamma_{xy,\text{cont}}$ and $\gamma_{yx,\text{cont}}$ quantify the instantaneous association encoded between the two variables in this representation.

In a continuous-time system, the state transitions over any finite interval $\Delta t$ are dictated by the matrix exponential of the underlying continuous-time drift matrix, $\mathbf{A}_{\text{cont}}$. 
Because the coefficients in Eqs.~(\ref{eq:disc-x})--(\ref{eq:disc-y}) are defined per iteration, the one-iteration transition matrix is $\mathbf{I}+\mathbf{B}_{\text{disc}}$:
\begin{equation}
    \mathbf{A}_{\text{disc}} = \begin{bmatrix}
    1 + \beta_{x,\text{disc}}
    & \gamma_{xy,\text{disc}} \\
    \gamma_{yx,\text{disc}}
    & 1 + \beta_{y,\text{disc}}
    \end{bmatrix}
\end{equation}
Under the time-homogeneous linear embedding described by Voelkle et al., the transition relation over interval $\Delta t$ is~\cite{voelkle2012sem}
\begin{equation}
    \mathbf{A}_{\text{disc}} = e^{\mathbf{A}_{\text{cont}} \cdot \Delta t}
\end{equation}
When the principal real matrix logarithm exists, the corresponding continuous-time drift follows by inversion of this relationship~\cite{voelkle2012sem}:
\begin{equation}
    \mathbf{A}_{\text{cont}} = \frac{\ln(\mathbf{A}_{\text{disc}})}{\Delta t} = \begin{bmatrix} \beta_{x,\text{cont}} & \gamma_{xy,\text{cont}} \\ \gamma_{yx,\text{cont}} & \beta_{y,\text{cont}} \end{bmatrix}
\end{equation}

For this diagonalizable matrix, we evaluate the matrix logarithm by eigendecomposition.
Let $\mathbf{A}_{\text{disc}} = \mathbf{V} \mathbf{D}_{\text{disc}} \mathbf{V}^{-1}$, where $\mathbf{D}_{\text{disc}} = \text{diag}(\lambda_1, \lambda_2)$ contains the discrete eigenvalues of the transition matrix~\footnote{Note: We use $\lambda_1, \lambda_2$ for the eigenvalues here to avoid a notation conflict with the latent slope multipliers $\mu_x, \mu_y$.} and $\mathbf{V}$ represents the matrix of corresponding eigenvectors. The continuous drift matrix is recovered by mapping the scalar natural logarithms of the eigenvalues back to the coordinate space:
\begin{equation}
    \mathbf{A}_{\text{cont}} = \frac{1}{\Delta t} \mathbf{V} \begin{bmatrix} \ln(\lambda_1) & 0 \\ 0 & \ln(\lambda_2) \end{bmatrix} \mathbf{V}^{-1}
\end{equation}

We then derive the affine continuous-time inputs $\mu_x$ and $\mu_y$ by matching the one-iteration affine transition.
Let $ \boldsymbol{\mu}_{\text{cont}} =[\mu_x, \mu_y]^T$. Then 
\begin{equation*}
\boldsymbol{\mu}_{\text{cont}} = \mathbf{M}^{-1}\boldsymbol{b}_{\text{disc}},
\qquad \boldsymbol{b}_{\text{disc}} = [\alpha_x s_x,\ \alpha_y s_y]^{T} .
\end{equation*}
where 
\begin{equation*}
\mathbf{M} \equiv \mathbf{A}_{\text{cont}}^{-1}
\left(e^{\mathbf{A}_{\text{cont}}\Delta t} - \mathbf{I}\right) .
\end{equation*}

We consider an illustrative 40-year trajectory from age 20 to age 60 with an annual step of $\Delta t=1$ year. Values are reported using the standard Wechsler IQ Scale Metric (mean = 100, SD = 15), while the endpoints are boundary conditions rather than population estimates. These conditions encode a linear decline in fluid intelligence from $X(20)=100$ to $X(60)=88$ and an increase in crystallized intelligence through education and experiential learning from $Y(20)=100$ to a later peak or plateau at $Y(60)=105$.

The deterministic parameterization satisfies three constraints: (1) Eqs.~(\ref{eq:par-beta})--(\ref{eq:par-mix}); (2) a stable-node fixed point, with stability conditions derived in Appendix~\ref{app:stability-fixed-point}; and (3) a relaxation time of 20--60 annual steps at $\Delta t=1$ year.
Appendix~\ref{app:feasible-ranges} outlines a sampling procedure for selecting model parameters.

One parameter set satisfying these constraints is $\beta_{x,\text{disc}} = -0.2086$, $\beta_{y,\text{disc}} = -0.1353$, $\gamma_{xy,\text{disc}} = 0.1026$, $\gamma_{yx,\text{disc}} = 0.1320$, $s_{x}=7.5660$, and $s_{y}=2.5677$, with $\alpha_x=\alpha_y=1$. Using the procedure outlined in this subsection the continuous-time system parameters are computed as: $\beta_{x,\text{cont}} = -0.2442$, $\beta_{y,\text{cont}} = -0.1551$, $\gamma_{xy,\text{cont}} = 0.1248$, $\gamma_{yx,\text{cont}} = 0.1606$, $\mu_x = 8.3692$, and $\mu_y = 2.1284$. Both systems, discrete and continuous-time, have the same stable equilibrium point at $X^{*}\approx 87.69$ and $Y^{*}\approx 104.52$. We use this stable human-only equilibrium as the model's biological baseline.

\subsection{Understanding AI Influence} \label{sub-sec:Understanding}

This subsection discusses how AI is hypothesized to influence human cognitive parameters. Because we analyze this influence in qualitative terms, we simplify the notation by using $\beta_x$, $\beta_y$, $\gamma_{xy}$, and $\gamma_{yx}$  for both discrete and continuous-time models. The positive and negative cases below define model scenarios, with $\omega_i$ treated as systematically varied parameters.

Coupling parameters describe how one intelligence domain contributes to the rate of change of another. The coefficient $\gamma_{yx}$ represents the investment relation in which fluid intelligence ($X$) contributes to the acquisition of crystallized knowledge ($Y$). A negative AI-mediated change in this coefficient represents weaker investment coupling: as $Z$ increases, a given level of $X$ contributes less to the accumulation of $Y$. One possible interpretation is that repeated offloading reduces the fluid reasoning exercised during knowledge acquisition. A positive change instead represents stronger investment coupling, as might occur if AI assistance reduces routine demands while preserving engagement with higher-order reasoning. These interpretations motivate the parameter ranges examined.

The coefficient $\gamma_{xy}$ represents the scaffolding relation in which crystallized knowledge ($Y$) supports fluid reasoning ($X$). A positive AI-mediated change in $\gamma_{xy}$ represents stronger scaffolding, which could occur if AI helps a user apply accumulated knowledge during novel problem solving. A negative change represents weaker scaffolding, as might occur if reliance on AI reduces the use of existing knowledge in unassisted reasoning. The self-feedback coefficient $\beta_x$ provides a separate channel: making $\beta_x$ more negative strengthens the damping of $X$, whereas making it less negative weakens that damping. Together, these coefficient changes define the model scenarios examined below.

The coefficient $\beta_y$ governs the proportional contribution of the current level of crystallized intelligence to its own rate of change. An AI-mediated change that makes $\beta_y$ more negative increases damping and lowers the equilibrium contribution associated with retaining crystallized knowledge. This may occur when external retrieval reduces the value of maintaining information internally. A change that makes $\beta_y$ less negative, or positive, instead represents weaker damping or self-reinforcing growth. This could model a tutoring scenario in which AI assistance supports the acquisition and retention of knowledge. As with the coupling coefficients, these interpretations connect parameter signs to the scenarios considered in the analysis.

\subsection{Modeling human-AI cognitive coevolution} \label{sec:modeling-human-AI}

The human-AI cognitive coevolutionary process is formalized as a three-dimensional coupled nonlinear dynamical system across two distinct timescales:
\begin{align}
    \frac{dX}{dt} &= F_x(X,Y; \mu_x, \beta_x, \beta_{xy}) + \tilde{\beta}_{x}X + \tilde{\beta}_{xy} Y \label{eq:slow-X}\\[1ex]
    \frac{dY}{dt} &=  F_y(X,Y; \mu_y, \beta_y, \beta_{yx}) + \tilde{\beta}_{y} Y + \tilde{\beta}_{yx} X \label{eq:slow-Y} \\[1ex]
\varepsilon\frac{dZ}{dt} &= Z \left(1 - \frac{Z}{Z_{\text{max}}  }\right) \label{eq:fast-Z}    
\end{align}
where $\varepsilon$ is a small perturbation parameter,  ($\varepsilon \to 0$), formalizing the separation of timescales. 
For the singular-perturbation analysis, we treat machine capability as fast relative to human cognitive change.

For simplicity, we write $\beta_x = \beta_{x,\text{cont}}$, $\beta_y = \beta_{y,\text{cont}}$, $\beta_{xy} = \gamma_{xy,\text{cont}}$, and $\beta_{yx} = \gamma_{yx,\text{cont}}$.
We define $\tilde{{\beta}}_{i}$, $i\in \{x,y,xy,yx \}$, so that $\tilde{{\beta}}_{i}=0$ when $Z=0$. Under this condition, Eqs.~(\ref{eq:slow-X})–(\ref{eq:slow-Y}) reduce to Eqs.~(\ref{eq:cont-x})–(\ref{eq:cont-y}). Thus, $F_x(X,Y; \mu_x, \beta_x, \beta_{xy})$ and $F_y(X,Y; \mu_y, \beta_y, \beta_{yx})$ denote the corresponding right-hand sides of Eqs.~(\ref{eq:cont-x})–(\ref{eq:cont-y}).

In this model, $\tilde{{\beta}}_{i}$, $i\in \{x,y,xy,yx \}$ are functions of the variables $X,Y$ and $Z$. These functions represent AI-mediated changes to the human cognitive coefficients. We choose a Hill function with $n=4$ to impose a delayed, sigmoidal transition between weak and saturated AI-mediated coefficient changes. Here, $Z$ represents composite machine capability, while $X$ and $Y$ represent fluid and crystallized intelligence. To make the Hill-function argument dimensionless, we compare $Z$ with the two-coordinate normalization $C$ and define
\begin{equation}  \label{eq:ratio}
R = \frac{Z}{C} \quad \text{where} \quad C = 100 + \frac{X + Y - 200}{\sqrt{2 + 2r}}
\end{equation}
Here $C$ is evaluated as outlined in Appendix~\ref{app:composite-score}.

Using this ratio, we model the AI-mediated change in each cognitive coefficient as
\begin{equation}
\begin{gathered}
\tilde{\beta}_{i}=\omega_i H(R),
\qquad H(R)=\frac{R^4}{R^4+1},\\
i\in\{x,y,xy,yx\}.
\end{gathered}
\label{eq:ai-coefficient-change}
\end{equation}
Here, $\omega_i$ sets the limiting magnitude and sign of the change in the corresponding coefficient as $H(R)$ approaches one. Specifically, $\omega_x$ and $\omega_y$ modify the self-feedback terms of $X$ and $Y$, respectively; $\omega_{xy}$ modifies the $Y\mathbin{\to}X$ coupling; and $\omega_{yx}$ modifies the $X\mathbin{\to}Y$ coupling. The resulting coefficient is
\begin{equation*}
\hat{\beta}_i=\beta_i+\omega_iH(R),
\end{equation*}
so $\omega_i=0$ recovers the calibrated human-only coefficient.

For fixed $X$ and $Y$ with $Z_{\max}>0$, Eq.~(\ref{eq:fast-Z}) is logistic in $Z$, with $Z(t)\to Z_{\max}$ as $t\to\infty$. We define the state-dependent carrying capacity as
\begin{equation} \label{eq:Z-max}
    Z_{\max} = K_0 + \kappa_{x} X + \kappa_{y} Y
\end{equation}
where $\kappa_x$ and $\kappa_y$ are phenomenological weights linking the human state variables to machine capability, and $K_0$ is an exogenous machine-capability term.

In Eq.~(\ref{eq:Z-max}), $K_0$ represents the component of the machine-capability ceiling that is not expressed as a function of the human state variables $X$ and $Y$. The model does not resolve the technical processes underlying this component separately. Recursive training on model-generated data can produce performance degradation under the conditions studied by Shumailov et al.~\cite{shumailov2024ai}. Other studies illustrate capabilities supported by particular architectures and training systems. MuZero uses learned, planning-relevant latent dynamics to support search in Go, chess, shogi, and Atari without a supplied perfect simulator~\cite{Schrittwieser2020MuZero}. AlphaFold predicts complex protein structures with atomic accuracy~\cite{jumper2021highly}. Deep-learning methods have also identified millions of candidate inorganic crystal structures, including 381,000 predicted to be thermodynamically stable~\cite{merchant2023scaling}. These studies illustrate heterogeneous technical influences on machine capability, but they do not provide an empirical estimate of $K_0$.

Since the parameters $K_0$, $\kappa_{x}$, and $\kappa_{y}$ are difficult to estimate, we construct a benchmark machine-capability value $Z_{\mathrm{ref}}$ from ten domain scores~\cite{hendrycks2025definitionagi}. Within this formulation, the ratio $R=Z/C$ requires the machine benchmark to be expressed on a reference scale compatible with the two human coordinates. Accordingly, we use the domain correspondence to calibrate $Z_{\mathrm{ref}}$ within the model's two-dimensional $X$--$Y$ state space. The ten domains are derived from CHC broad abilities and weighted equally at 10\%. To connect the ten machine domains to the two human state variables, we assign each included domain to either the fluid or crystallized component. CHC theory treats several of the corresponding human abilities as broad domains distinct from both $G_f$ and $G_c$~\cite{mcgrew2009chc}, and we group them according to their role in the two-state model. We assign On-the-Spot Reasoning (R), Working Memory (WM), and Processing Speed (S) to the fluid component and General Knowledge (K), Reading and Writing (RW), Mathematical Ability (M), Long-Term Memory Storage (MS), and Long-Term Memory Retrieval (MR) to the crystallized component:
\begin{align}
G_f &\;\longleftrightarrow\; \text{R} + \text{WM} + \text{S} ,
\label{eq:assign-gf}\\
G_c &\;\longleftrightarrow\; \text{K} + \text{RW} + \text{M} + \text{MS} + \text{MR} .
\label{eq:assign-gc}
\end{align}
Visual (V) and auditory (A) processing are excluded because the Cattell-Horn-Carroll hierarchy treats $G_v$ and $G_a$ as Stratum II broad abilities distinct from both $G_f$ and $G_c$.  
Appendix~\ref{app:cognitive_mapping} states the modeling rationale for each assignment and quantifies sensitivity to alternative reductions.

Using the frontier-model scores~\cite{hendrycks2025definitionagi}, we express each group as a percentage of its own maximum because the two groups contain different numbers of domains:
\begin{align}
X_{AI} &= \tfrac{7+4+3}{30}\times 100 = 46.7 , \label{eq:xai}\\
Y_{AI} &= \tfrac{9+10+10+0+4}{50}\times 100 = 66.0 . \label{eq:yai}
\end{align}
Passing this pair through the composite of Eq.~(\ref{eq:ratio}) with $r=0.65$, so that $\sqrt{2+2r} = 1.8166$, and doing the same for the unaugmented human baseline $(X^{*}, Y^{*}) = (87.68,\, 104.52)$ of
Sec.~\ref{sec:model_formulation},
\begin{align}
Z_{\mathrm{ref}} &= 100 + \frac{46.7 + 66.0 - 200}{1.8166} = 51.92 ,
\label{eq:zmax-value}\\
C^{*} &= 100 + \frac{87.68 + 104.52 - 200}{1.8166} = 95.70 ,
\label{eq:cstar-value}
\end{align}
so that the benchmark ratio and the corresponding value of the Hill term are
\begin{align}
R_{\mathrm{ref}} &= \frac{Z_{\mathrm{ref}}}{C^{*}} = 0.543 , \label{eq:rtoday}\\
H(R_{\mathrm{ref}}) &= \frac{R_{\mathrm{ref}}^{4}}{R_{\mathrm{ref}}^{4}+1} = 0.080 .
\label{eq:htoday}
\end{align}
Under the adopted domain assignment and composite normalization, the frontier-model scores yield the constructed benchmark ratio $R_{\mathrm{ref}}=0.543$, or roughly $54\%$ relative to the calibrated human composite $C^*=95.70$, with $H(R_{\mathrm{ref}})=0.080$. Thus, the Hill modifier is approximately $8\%$ of its limiting value at the reference point. This is a model calibration rather than a validated psychometric comparison of human and machine intelligence. The resulting value is numerically close to the overall capability score of $57\%$~\cite{hendrycks2025definitionagi}, but it is obtained differently: that score weights all ten domains equally, whereas Eqs.~(\ref{eq:assign-gf})--(\ref{eq:rtoday}) pass the two-group reduction through the composite of Appendix~\ref{app:composite-score}.

To express this same benchmark through the state-dependent carrying capacity in Eq.~(\ref{eq:Z-max}), we take the relative weights to follow $X_{AI}:Y_{AI}$ and choose their common scale so that $Z_{\max}(X^*,Y^*;0)=Z_{\mathrm{ref}}$. This gives
\begin{align*}
(\kappa_x,\kappa_y)&=(0.220494,\,0.311842),\\
\kappa_xX^*+\kappa_yY^*&=51.9246=Z_{\mathrm{ref}}.
\end{align*}
The numerical simulations use the same reference values $R_{\mathrm{ref}}=0.5426$ and $H(R_{\mathrm{ref}})=0.07975$ while varying $K_0$. The scale is criterion-referenced, with $100$ denoting the performance of a well-educated adult rather than a population standard score~\cite{hendrycks2025definitionagi}. We use that endpoint as the state-dependent carrying-capacity cap, $Z_{\text{max}}=100$, corresponding to $K_0\leq48.08$ when $X$ and $Y$ are fixed at the baseline. Because $Z_{\max}=K_0+\kappa_xX+\kappa_yY$ is state-dependent, $K_0=48.08$ is only the fixed-reference equivalent of this bound, not a universal ceiling on $K_0$; the analysis is bounded at $Z_{\max}=100$ for each evaluated state. We record as limitations both the extrapolation beyond this criterion-referenced scale and the mismatch between its well-educated-adult reference and the mean-$100$ human presentation scale. Appendix~\ref{app:fixed-point} contains derivation of these fixed points.

\section{AI-induced Cognitive  Augmentation / Decline}
\label{sec:augmentation-outcomes}

The model \eqref{eq:slow-X}--\eqref{eq:fast-Z} can capture both the positive and negative effects of AI on human cognition. Let $(\hat{X}, \hat{Y}, \hat{Z})$ denote a steady-state solution of the coupled system \eqref{eq:slow-X}--\eqref{eq:fast-Z}. We ask under what conditions AI-mediated changes increase or decrease one or both human capacities. To determine these conditions, we compare each augmented equilibrium with a common reference point. We use the stable equilibrium of Eqs.~(\ref{eq:cont-x})--(\ref{eq:cont-y}) before any AI-mediated coefficient changes, $(X^*,Y^*)=(87.68,104.52)$, and refer to it as the biological baseline. The steady-state changes are then defined as
\begin{equation}
\Delta X=\hat X-X^*, \qquad \Delta Y=\hat Y-Y^*.
\label{eq:delta-def}
\end{equation}
Thus, AI-induced augmentation requires $\Delta X>0$ and $\Delta Y>0$,
AI-induced decline requires both changes to be negative, and a mixed outcome has one change of each sign.

\begin{figure*}[!htbp]
\centering
\includegraphics[width=\textwidth]{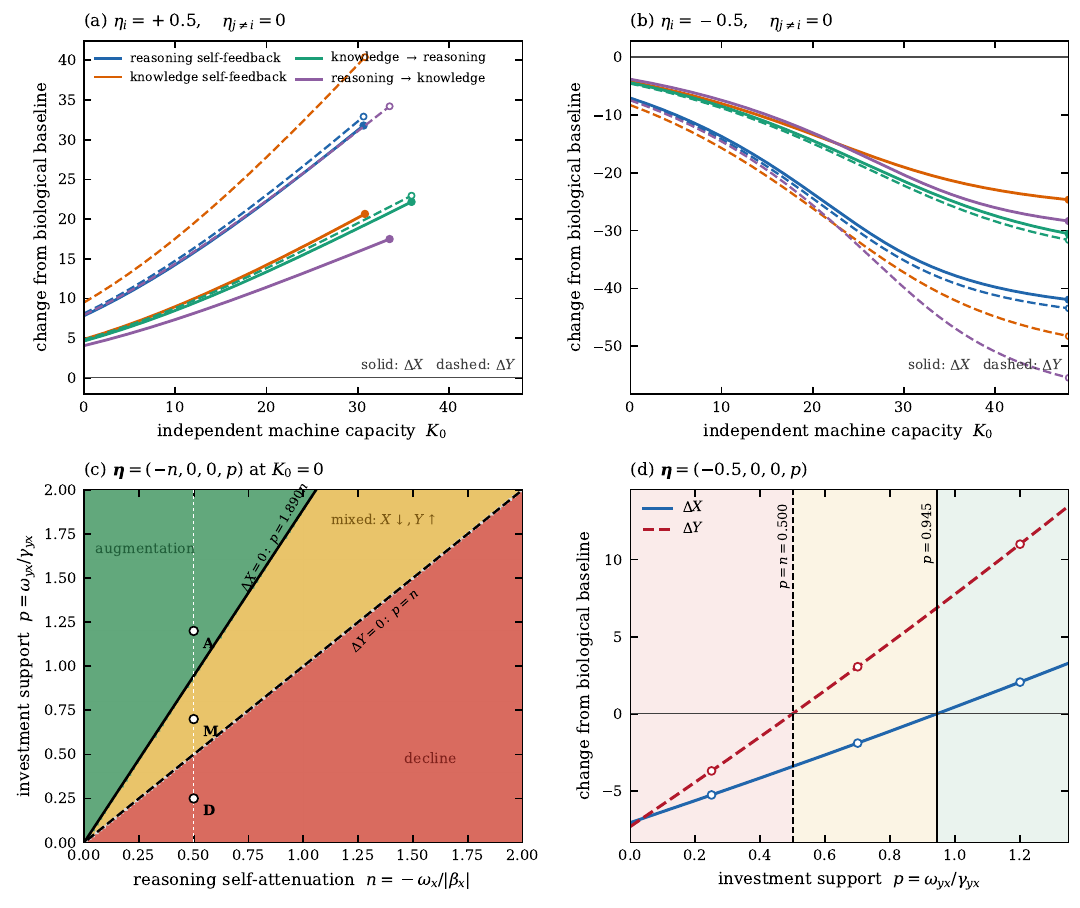}
\caption{Steady-state changes defined in Eq.~(\ref{eq:delta-def}) relative to the biological baseline $(X^*,Y^*)=(87.68,104.52)$. With
$\boldsymbol{\eta}=(\eta_x,\eta_y,\eta_{xy},\eta_{yx})$ defined by
Eq.~(\ref{eq:omega-normalization}), panel~(a) uses the four separate settings $(+0.5,0,0,0)$, $(0,+0.5,0,0)$, $(0,0,+0.5,0)$, and
$(0,0,0,+0.5)$. Panel~(b) uses the corresponding four settings with $-0.5$. Thus, only one normalized parameter is nonzero in each experiment. The legend identifies that parameter: $\eta_x$ modifies the $X$ self-term, $\eta_y$ the $Y$ self-term, $\eta_{xy}$ the $Y\mathbin{\to}X$ coupling, and $\eta_{yx}$ the $X\mathbin{\to}Y$ coupling. Each color corresponds to two curves: solid for $\Delta X$ and dashed for $\Delta Y$. Panels~(c) and~(d) instead use $\boldsymbol{\eta}=(-n,0,0,p)$ at $K_0=0$, where $n=-\omega_x/|\beta_x|$ and $p=\omega_{yx}/\gamma_{yx}$. In panel~(c), red denotes decline, yellow the mixed outcome $\Delta X<0<\Delta Y$, and green augmentation. The solid and dashed boundaries mark $\Delta X=0$ and $\Delta Y=0$, respectively. Panel~(d) fixes $n=0.5$; $Y$ returns to baseline at $p=0.500$ and full augmentation begins at $p=0.945$. All curves follow the stable equilibrium connected continuously to the biological baseline and are displayed only while $40\leq X,Y\leq145$ and $Z_{\max}\leq100$.}
\label{fig:augmentation-outcomes}
\end{figure*}

In the model, the four AI-mediated changes act on coefficients of different magnitudes. We therefore compare them in the normalized form:
\begin{equation}
\boldsymbol{\eta}=
\left(
\frac{\omega_x}{|\beta_x|},\,
\frac{\omega_y}{|\beta_y|},\,
\frac{\omega_{xy}}{\gamma_{xy}},\,
\frac{\omega_{yx}}{\gamma_{yx}}
\right).
\label{eq:omega-normalization}
\end{equation}
A normalized value of $+0.5$, for example, means that the
largest change permitted by the Hill term is one half of the corresponding baseline coefficient. The realized change is smaller by the factor $H(R)$.
Panels~(a) and~(b) of Fig.~\ref{fig:augmentation-outcomes} vary one component of $\boldsymbol{\eta}$ at a time and set the other three to zero. Panel~(a) uses the four settings $(+0.5,0,0,0)$, $(0,+0.5,0,0)$,
$(0,0,+0.5,0)$, and $(0,0,0,+0.5)$; panel~(b) uses the corresponding
settings with $-0.5$. In panels~(a) and~(b), every plotted equilibrium has $\Delta X,\Delta Y>0$ for the four $+0.5$ parameter variations and $\Delta X,\Delta Y<0$ for the four $-0.5$ variations. Although only one coefficient is changed in each experiment,
$X$ and $Y$ remain positively coupled: $Y$ enters the $X$ equation through $\gamma_{xy}Y$, and $X$ enters the $Y$ equation through $\gamma_{yx}X$, with both coupling coefficients positive. Consequently, a coefficient variation that increases one equilibrium capacity also increases the other, while a variation that decreases one capacity also decreases the other. Hence, no mixed outcome occurs when the normalized parameters are varied individually.

Panels~(a) and~(b) also show that increasing $K_0$ increases the absolute equilibrium changes $|\Delta X|$ and $|\Delta Y|$.
From $Z_{\max}=K_0+\kappa_xX+\kappa_yY$, a larger $K_0$ raises $R$
and hence $H(R)$ along these trajectories. A larger $H(R)$ makes the same normalized parameter change produce a larger realized change $\omega_iH(R)$. At $K_0=0$, $Z_{\max}=\kappa_xX+\kappa_yY$ is still positive because it is supported by the human capacities. Consequently, the equilibria at $K_0=0$ already differ from the biological baseline.

The results in panels~(a) and~(b) show that changing any single coefficient moves $X$ and $Y$ in the same direction. This does not address situations in which AI weakens one human feedback process while strengthening another. We therefore ask which effect dominates and whether their balance produces decline, a mixed outcome, or augmentation. To examine this question, we combine $\omega_x<0$, which adds a negative term to the $X$ equation, with
$\omega_{yx}>0$, which increases the contribution of $X$ to the $Y$
equation:
\begin{equation}
n=-\frac{\omega_x}{|\beta_x|}\geq0,\qquad
p=\frac{\omega_{yx}}{\gamma_{yx}}\geq0,
\qquad \omega_y=\omega_{xy}=0.
\label{eq:competition-slice}
\end{equation}
Here $n$ changes the $X$ self-term through
$\omega_x=-n|\beta_x|$, while $p$ changes the $X\mathbin{\to}Y$ coupling through $\omega_{yx}=p\gamma_{yx}$. Subtracting the two biological equilibrium equations from the augmented equilibrium equations gives
\begin{align}
0
&{=\beta_x\Delta X+\gamma_{xy}\Delta Y
-n|\beta_x|H(\hat R)\hat X,} \nonumber\\
{0}
&{=\gamma_{yx}\Delta X+\beta_y\Delta Y
+p\gamma_{yx}H(\hat R)\hat X.}
\label{eq:competition-deltas}
\end{align}
Setting $\Delta Y=0$ in these equations gives $p=n$. Setting $\Delta X=0$ gives a second straight boundary,
\begin{equation}
\begin{aligned}
\Delta Y=0 &: \quad p=n,\\
\Delta X=0 &: \quad p=Dn,\\
D&=\frac{|\beta_x\beta_y|}{\gamma_{xy}\gamma_{yx}}=1.8897.
\end{aligned}
\label{eq:augmentation-boundaries}
\end{equation}
The common factors $H(\hat R)$ and $\hat X$ cancel. The slopes therefore do not depend on $K_0$, $\kappa_x$, $\kappa_y$, or the Hill exponent. For the calibrated coefficients, the slice separates into three regions.
\begin{equation}
\begin{aligned}
p<n &:\quad \Delta X<0,\ \Delta Y<0,\quad \text{decline},\\
n<p<1.8897n &:\quad \Delta X<0,\ \Delta Y>0,\quad \text{mixed},\\
p>1.8897n &:\quad \Delta X>0,\ \Delta Y>0,\quad \text{augmentation}.
\end{aligned}
\label{eq:augmentation-regions}
\end{equation}
Increasing $n$ strengthens the negative effect on fluid reasoning. This lowers $X$ directly and, because $X$ contributes to knowledge accumulation, also places downward pressure on $Y$. Increasing $p$, by contrast, strengthens the contribution of fluid reasoning to crystallized knowledge. This raises $Y$, which in turn supports $X$ through the positive $Y\mathbin{\to}X$ coupling. When $p<n$, this positive contribution is too weak to counteract the stronger damping of $X$, and both capacities remain below baseline. When $n<p<1.8897n$, knowledge rises above baseline, but the indirect support from $Y$ is not yet sufficient to restore fluid reasoning. This gives a mixed outcome. Once $p>1.8897n$, the feedback from the higher level of $Y$ is strong enough to raise $X$ above baseline, and both capacities exceed baseline.

The opposite mixed outcome, $\Delta X>0>\Delta Y$, is absent
from this two-parameter quadrant. It arises under the complementary choice \(\omega_y<0\) and \(\omega_{xy}>0\), with \(\omega_x=\omega_{yx}=0\), which produces the analogous progression with the roles of \(X\) and \(Y\) reversed. At $n=0.5$, knowledge first returns to its biological level at $p=0.5$, but fluid reasoning remains below baseline until $p=0.945$. The three marked equilibria illustrate the progression. At $p=0.25$, $(\Delta X,\Delta Y)=(-5.80,-4.08)$; at $p=0.70$, the outcome is mixed with $(-2.10,+3.36)$; and at $p=1.20$, both coordinates increase, with
$(+2.27,+12.18)$. A negative effect on fluid reasoning does not necessarily cause both capacities to decline. A sufficiently large $\omega_{yx}>0$ raises $Y$ enough that its feedback lifts both $X$ and $Y$ above baseline.

\section{Bifurcation analysis} \label{sec:results}

The preceding analysis classifies how AI-mediated coefficient changes shift equilibria relative to the biological baseline. We now examine how those changes alter their existence and stability. The human-AI cognitive coevolution model comprises two subsystems: a fast technology subsystem and a slow human cognitive subsystem. In the absence of AI, Eqs.~(\ref{eq:slow-X})–(\ref{eq:slow-Y}) reduce to Eqs.~(\ref{eq:cont-x})–(\ref{eq:cont-y}). The slow--fast reduction based on Tikhonov's theorem and the derivation of the saddle-node condition are provided in Appendix~\ref{app:tikhonov}.

Throughout the bifurcation analysis, the AI-mediated coefficient changes follow Eq.~(\ref{eq:ai-coefficient-change}). For any given value of $R$, the change needed to cancel the original coefficient is
\begin{equation}
\begin{aligned}
\omega_i^{\text{flip}} &= -\frac{\beta_i}{H(R)},
|\omega_i^{\text{flip}}| &= \frac{|\beta_i|}{H(R)} \;\geq\; |\beta_i|,
\qquad H \in (0,1) .
\end{aligned}
\label{eq:omega-flip}
\end{equation}

The coupled system can exhibit several dynamical behaviors. Here we restrict the analysis to two illustrative cases: saddle-node and Hopf bifurcations.

\subsection{Saddle-node Bifurcation} \label{sec:caseI}

We next examine how an AI-induced weakening of the cross-couplings between $X$ and $Y$ affects the stable equilibrium. A negative $\omega _{yx}$ attenuates the contribution of fluid reasoning to knowledge accumulation, whereas a negative $\omega _{xy}$ weakens the effect of accumulated knowledge on fluid reasoning. The parameter $K_{0}$ represents an independent machine capacity. Increasing $K_{0}$ raises $R$ and $H(R)$ across the states examined here, thereby amplifying the dynamics specified by $\omega _{i}$.

When AI weakens the contribution of fluid reasoning to
knowledge accumulation ($\omega_{yx}<0$), increasing $K_0$ progressively reduces the support that $X$ provides to $Y$. Knowledge settles at a lower equilibrium level because $X$ provides less support. The resulting decline in $Y$ weakens its positive contribution to $X$, so fluid reasoning also declines. For $\omega_{yx}=-1.5\,\gamma_{yx}$, the stable fixed point and the
saddle coexist for $8.84\leq K_0\leq 11.98$ and meet at
$K_0^{*}=11.98$ in a saddle-node bifurcation. Beyond this point, no fixed point with positive $Y$ remains (Fig.~\ref{fig:phase}, panel~(a)). The fold calculation and parameter values are given in
Appendices~\ref{app:tikhonov} and~\ref{app:caseI-details}.

To compare the two relations, we first vary one parameter at a
time. When only the contribution of $X$ to $Y$ is weakened
($\omega_{yx}<0$, with $\omega_{xy}=0$), the first saddle-node
appears at
\begin{equation}
\omega_{yx} = -1.234\,\gamma_{yx} .
\label{eq:threshold-fold}
\end{equation}
A larger change, $\omega_{xy}=-2.605\,\gamma_{xy}$, is required when only the contribution of $Y$ to $X$ is weakened ($\omega_{xy}<0$, with $\omega_{yx}=0$). A saddle-node therefore occurs after a smaller proportional weakening of the contribution of $X$ to $Y$ than of the reverse contribution from $Y$ to $X$ (Fig.~\ref{fig:foldmap}, panel~(b)). This asymmetry follows from the calibration, in which fluid reasoning supplies most of the input that maintains accumulated knowledge.

Panel~(b) also shows the joint effect of the two parameters. Combinations in the pale region produce no admissible fold before $Z_{\max}=100$; below the onset boundary, the color gives the value of $K_0$ at which the stable fixed point disappears. The two parameters affect the onset and location of a fold differently. Changing $\omega_{xy}$ from zero to $-2\gamma_{xy}$ moves the investment threshold from $-1.234\,\gamma_{yx}$ to $-1.427\,\gamma_{yx}$. Thus, weakening the contribution of $Y$ to $X$ can require a larger weakening of the contribution of $X$ to $Y$ before a fold first appears. Once a fold exists, however, the same change can make it occur at a lower $K_0$. At $\omega_{yx}=-1.5\,\gamma_{yx}$, for example, changing $\omega_{xy}$ from zero to $-\gamma_{xy}$ moves the fold from $K_0^{*}=11.98$ to $K_0^{*}=9.06$.

The numerical threshold also depends on the constant input $\mu_y$. When more of the maintenance of $Y$ is assigned to $\mu_y$, weakening the contribution from $X$ removes a smaller share of its total support. Appendix~\ref{app:caseI-details} reports how this assumption changes the threshold.

\begin{figure}[!htbp]
\centering
\includegraphics[width=0.95\textwidth]{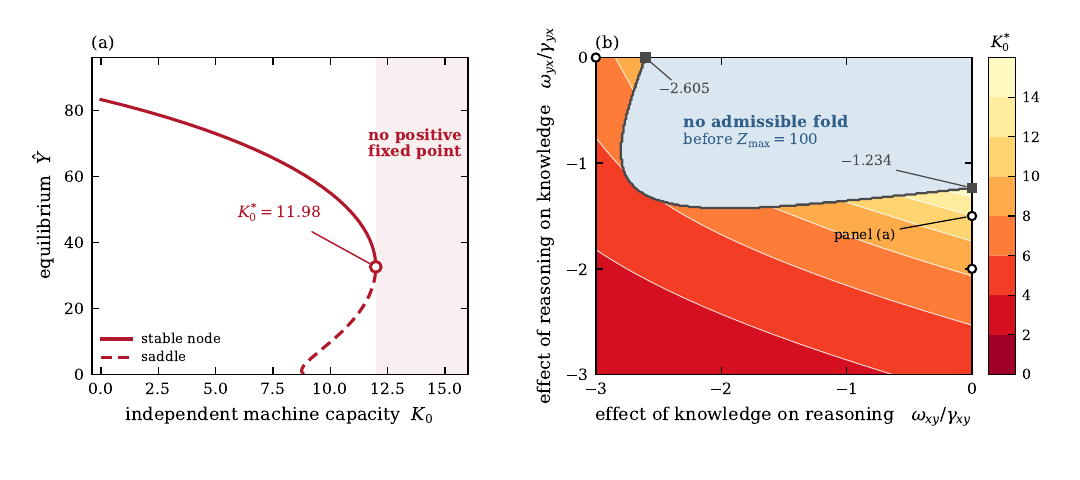}
\caption{{Saddle-node structure under AI-related weakening of the cross-couplings. The normalizations use the calibrated continuous-time coefficients $\gamma_{xy}=0.1248$ and $\gamma_{yx}=0.1606$; in both panels, $\omega_x=\omega_y=0$, and all remaining parameters retain their calibrated values. (a) Equilibrium crystallized intelligence $\hat{Y}$ as $K_0$ increases for $\omega_{yx}=-1.5\,\gamma_{yx}$ and $\omega_{xy}=0$. The solid stable-node branch meets the dashed saddle branch at $K_0^{*}=11.98$; the shaded region has no positive fixed point. (b) Joint action of $\omega_{xy}$ and $\omega_{yx}$. Moving left weakens $Y\to X$, and moving down weakens $X\to Y$. Color gives the $K_0$ at which the stable state disappears. The pale region has no admissible fold before $Z_{\max}=100$,} and the heavy curve marks the onset boundary.}
\label{fig:phase}
\label{fig:foldmap}
\end{figure}

The saddle-node also affects recovery from small disturbances before the stable fixed point disappears. If $\delta=K_0^{*}-K_0$, the recovery rate follows
\begin{equation}
{|\mathrm{Re}\,\lambda_{\text{slow}}| \sim \delta^{\,\text{0.462}}, \qquad R^{2} = 0.999} ,
\label{eq:slowing}
\end{equation}
close to the square-root scaling expected near a saddle-node. As $K_0$ approaches $K_0^{*}$, recovery from a disturbance to $X$ or $Y$ becomes progressively slower. The calculation is reported in
Appendix~\ref{app:caseI-details}.

We also examined whether the formulation could produce limit cycles instead of settling at a fixed point. No limit cycles were found for $40\leq X,Y\leq140$, the score range examined in this search (Appendix~\ref{app:caseI-details}). Section~\ref{sec:hopf} considers a different combination of parameter changes that produces this behavior.

\subsection{Hopf Bifurcation}
\label{sec:hopf}

The preceding results concern parameter changes for which fluid reasoning $X$ and crystallized knowledge $Y$ settle to constant levels or eventually leave the range represented by the model. We next consider a different combination of AI effects under which fluid reasoning and crystallized knowledge vary periodically. We examine this regime as a structural possibility within the model.

In the baseline model, fluid reasoning and crystallized knowledge support one another. Fluid reasoning contributes to the accumulation of crystallized knowledge, which in turn supports fluid reasoning. This mutually supportive relation does not provide the opposing feedback needed for a repeating cycle. Such a cycle becomes possible only when AI reverses one direction of the relation. A sufficiently
negative $\omega_{xy}$ reverses the net $Y\mathbin{\to}X$ coupling, so that $\hat\beta_{xy}<0$, while the $X\mathbin{\to}Y$ coupling remains positive ($\hat\beta_{yx}>0$). In this parameter regime, AI mediation changes the modeled role of crystallized knowledge: instead of supporting novel reasoning, its net contribution becomes negative. Under one interpretation, AI changes how accumulated knowledge is used, making it more likely to substitute for novel reasoning. We represent this possibility phenomenologically through the sign reversal.

This change creates an opposing feedback loop between the two cognitive capacities. A temporary increase in fluid reasoning
raises crystallized knowledge through the positive
$X\mathbin{\to}Y$ coupling. Because the effective $Y\mathbin{\to}X$ coupling is negative, the subsequent increase in $Y$ pushes $X$ downward. Lower $X$ then provides less support for knowledge accumulation, so $Y$ falls and its negative contribution to $X$ weakens. Below the Hopf point, these alternating deviations become progressively smaller and the joint state $(X,Y)$ returns to
its equilibrium. Increasing $\omega_x$ strengthens the self-feedback of fluid reasoning and reduces this damping. At the Hopf point, the equilibrium loses stability. Above it, small deviations grow initially, but nonlinear feedback limits their amplitude and the trajectory approaches a stable periodic orbit.

\begin{figure}[!htbp]
\centering
\includegraphics{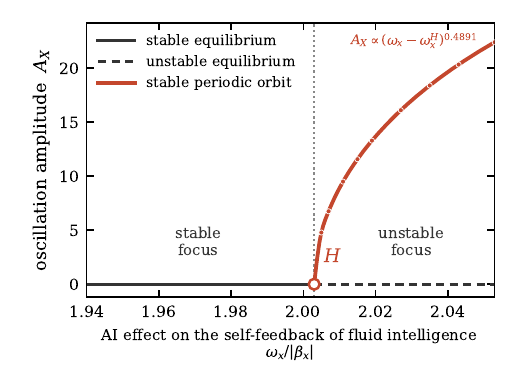}
\caption{One-parameter diagram of the supercritical Hopf bifurcation. We vary the AI effect on fluid-reasoning self-feedback $\omega_x$, holding $\omega_{xy}=-3.814\,\gamma_{xy}$, $\omega_y=-0.603\,|\beta_y|$, $\omega_{yx}=+1.100\,\gamma_{yx}$, and $K_0=42.1$ fixed. The normalizations use the calibrated continuous-time coefficients $\beta_x=-0.2442$, $\beta_y=-0.1551$,
$\gamma_{xy}=0.1248$, and $\gamma_{yx}=0.1606$; all remaining parameters retain their calibrated values. At $H$ ($\omega_x^H=2.0030\,|\beta_x|$), the stable equilibrium (solid black) becomes unstable (dashed black) and a stable periodic
orbit (orange) emerges continuously. Its plotted amplitude is
$A_X=(X_{\max}-X_{\min})/2$; the measured exponent $0.4891$ is close to the square-root scaling of the Hopf normal form.}
\label{fig:hopf}
\end{figure}

To trace this transition in one parameter, we fix $\omega_{xy}=-3.814\,\gamma_{xy}$, a representative value from the
numerical search for which varying $\omega_x$ crosses a Hopf point and produces an attracting cycle. The cycle emerges continuously as $\omega_x$ increases, with amplitude exponent $0.4891$
($R^2=0.9999$), consistent with a supercritical Hopf bifurcation. At
$\omega_x=2.030\,|\beta_x|$ and $\omega_{xy}=-3.814\,\gamma_{xy}$, trajectories starting from different initial conditions converge to the same cycle. One cycle takes $31.5$ yr; during it, $X$ ranges from $8.6$ to $42.6$ and $Y$ from $25.9$ to $60.9$. Figure~\ref{fig:hopf} shows the onset as a one-parameter bifurcation diagram. Appendix~\ref{app:hopf-details} reports the numerical search, continuation, and coefficient-level checks.

\section{Discussion and Conclusions} \label{sec:disc-and-conc}

We developed a minimal, phenomenological model of human–AI cognitive coevolution by coupling a fast machine-capability variable to fluid and crystallized intelligence through Hill-function modifiers of the human self-feedback and cross-coupling coefficients. Relative to the biological baseline, the model produces decline, augmentation, or a mixed outcome in which one cognitive capacity rises while the other falls (Sec.~\ref{sec:augmentation-outcomes}). These outcomes depend on the balance between changes in fluid self-feedback and investment coupling into crystallized knowledge. On the slow manifold obtained through the Tikhonov reduction, the system also exhibits two qualitatively distinct losses of stability. Sufficient weakening of the fluid-to-crystallized coupling produces a saddle-node bifurcation (Sec.~\ref{sec:caseI}), whereas reversal of the crystallized-to-fluid coupling, followed by reduced damping of fluid reasoning, produces a supercritical Hopf bifurcation (Sec.~\ref{sec:hopf}). Within the model, short-term task performance does not determine the long-run dynamical regime.

Evidence from specific digital interventions shows that their design can shape cognitive outcomes. Action video games have been associated with enhanced visual selective attention~\cite{green2003action} and 
cognitive control, sustained attention, and working memory in older adults~\cite{anguera2013video}. In a randomized field experiment involving nearly 1,000 Turkish high-school students, unrestricted GPT-4 assistance increased mathematics-practice performance by approximately 48\%, but students subsequently scored 17\% below the no-AI control group on a closed-book examination. A guardrailed GPT-4 tutor produced a larger assisted-practice gain without a significant deficit on the later examination~\cite{bastani2025generative}. The present framework complements this line of work by providing a dynamical perspective on longer-term cognitive change. Within the model, avoiding the decline regime identified in Sec.~\ref{sec:caseI} requires coefficient changes that preserve or strengthen the active contribution of human reasoning to knowledge acquisition. Whether particular deployment strategies produce such changes remains an empirical question. The remainder of this section examines the robustness and limitations of the model, outlines directions for future work, and closes with a more speculative phenomenological reading of the coupled socio-technical system beyond the parameter ranges examined above.

Three limitations bound the interpretation of these results. 
First, the model uses a cross-sectional benchmark of frontier-model capability and an illustrative normal-aging parameterization (Sec.~\ref{sec:model_formulation}), not an empirically fitted longitudinal record of human cognitive change under sustained AI use. We are not aware of such a multi-year record, and longitudinal AI-cognition evidence remains scarce~\cite{znidaric2026review}.
The parameters $\omega_i$ governing AI's influence on human coupling coefficients are systematically varied rather than fitted, so the saddle-node and Hopf thresholds are conditional on the parameter ranges and functional form examined.

Second, the model is deterministic and two-dimensional on the slow manifold. It does not represent heterogeneity across individuals or professions, nor stochastic fluctuation around the mean trajectory, both of which would be needed before the critical slowing reported in Sec.~\ref{sec:caseI} could be treated as an operational monitoring target rather than a theoretical prediction.

Third, the composite metric $C$ (Appendix~\ref{app:composite-score}) and the assignment of domains to the two components (Appendix~\ref{app:cognitive_mapping}) rest on the assumption that machine and human cognitive capability are commensurable on a shared psychometric scale. This assumption makes $R$ well defined for the analysis. Appendix~\ref{app:cognitive_mapping} examines the sensitivity of this mapping on the human side. Assigning the sensory domains (V and A) in Hendrycks et al.'s~\cite{hendrycks2025definitionagi} ten machine domains to either component changes $R_{\text{ref}}$ only slightly. McGrew describes the older combined long-term storage and retrieval factor $G_{lr}$ as a broad ability distinct from $G_f$ and $G_c$~\cite{mcgrew2009chc}. We group Hendrycks et al.'s separate MS and MR domains under $G_{lr}$ and assign them to the crystallized component; this step materially affects the benchmark.

Several extensions would test the assumptions of the present model. Longitudinal cohort data from AI-assisted professional or student populations could be used to estimate the $\omega_i$ and convert the qualitative bifurcation map in Fig.~\ref{fig:foldmap} into a data-calibrated map. Allowing $\varepsilon$ to vary would extend the fast-AI limit $\varepsilon \to 0$ used in Sec.~\ref{sec:results} toward the regime described with $\varepsilon=O(1)$, where human and machine capability evolve on comparable timescales. This extension would test whether the identified fold is delayed or avoided outside the singular limit. Dividing the crystallized-intelligence state variable into domain-specific components, such as procedural and declarative knowledge, would test whether the asymmetry between the investment and scaffolding channels in Sec.~\ref{sec:caseI} persists at a finer resolution. A population model could couple individual $(X,Y)$ trajectories through shared institutional or educational structures and connect the analysis to the population-level deskilling evidence reviewed in Appendix~\ref{sec:appendix_evidence_offloading}. Network-based, spatially distributed agent models or stochastic Langevin formulations could then represent heterogeneity and fluctuations that are absent from the present single-agent system.

Section~\ref{sec:results} reports bifurcations within the parameter ranges examined. Here we consider a qualitative extension of the same structure beyond those ranges. The interpretation retains a fast machine-intelligence variable with a state-dependent carrying capacity coupled to a slow human cognitive subsystem, but it does not constitute an additional bifurcation result.

Because $Z_{\text{max}}$ is state-dependent through Eq.~\ref{eq:Z-max}, AI expansion can accompany either direction of steady-state change defined in Sec.~\ref{sec:augmentation-outcomes}. In the decline scenario, the coefficient changes accelerate the loss of fluid intelligence ($X$) and weaken investment into crystallized intelligence ($Y$), as in the parameter regime examined in Sec.~\ref{sec:caseI}. In the augmentation scenario, the coefficient changes raise both $X$ and $Y$ at equilibrium, as in the mixed-to-augmentation region of Fig.~\ref{fig:augmentation-outcomes}(c)--(d). These outcomes are sign patterns $(\Delta X,\Delta Y)$ generated by the specified $\omega_i$ combinations. Psychological accounts, such as automation freeing capacity for higher-order reasoning, are interpretations of those parameter regimes rather than results of the model.

Continuation beyond the calibrated horizon gives two hypothetical late trajectories. In the decline scenario, $X(t)$ and $Y(t)$ move away from the biological baseline after the saddle-node loss of the stable equilibrium described in Sec.~\ref{sec:caseI}. In the augmentation scenario, the coupled system approaches a new stable equilibrium. These continuations are qualitative illustrations rather than predictions derived from the calibrated analysis.

The analysis in Sec.~\ref{sec:results} fixes the time-scale separation parameter at the singular limit $\varepsilon \to 0$. In this limit, $Z$ rapidly approaches $Z_{\text{max}}$, and the reduced planar system in Appendix~\ref{app:tikhonov} governs the slow dynamics. Future work can examine the $\varepsilon=O(1)$ regime. When human and machine capability evolve on comparable timescales, the human subsystem may adjust before $Z$ reaches saturation, which could change whether the saddle-node found in Sec.~\ref{sec:caseI} is reached. This possibility has not been tested in the present analysis.

Taken together, the model produces conditional outcomes that include decline, augmentation, equilibrium loss, and oscillatory behavior. The resulting regime depends on how AI-mediated changes reshape self-feedback and cross-coupling between fluid reasoning and crystallized knowledge. This dependence is the central structural result of the analysis.

\begin{acknowledgments}
This research was partially supported by the Ministry of Education and Science through the project ``Utilizing AI and National Large Language Models to Advance Macedonian Language Capabilities,'' by the European Union under the Grant Agreement 101263128, by the Brain Science and Brain-like Intelligence Technology, National Science and Technology Major Project (Grant No. 2025ZD0215700), by the 
National Natural Science Foundation of China (Grant No 62441614), and by the ANSO, project number ANSO-CR-PP-2022-05. 
\end{acknowledgments}

\appendix

\section{Evidence for Offloading and Deskilling} \label{sec:appendix_evidence_offloading}

This appendix reviews key literature on AI-driven cognitive offloading and deskilling. Cross-sectional evidence associates heavier AI use with greater cognitive offloading and lower self-reported or measured critical-thinking outcomes, but it does not establish skill decline~\cite{gerlich2025aitools}.
In a cross-sectional survey of knowledge workers, greater task-specific confidence in GenAI was associated with less reported critical-thinking activity, while greater trust was associated with lower perceived effort for several cognitive activities~\cite{lee2025}.
Among Chinese university students, greater self-reported AI dependence was associated with lower self-reported critical thinking and greater cognitive fatigue; fatigue statistically mediated part of the association, and information literacy buffered rather than amplified the adverse associations in the reported interaction models~\cite{tian2025learners}. The remaining studies in this appendix use a mixture of experimental, observational, and conceptual designs and are interpreted according to those designs.
Automation complacency leads human operators to accept erroneous outputs uncritically, especially under low individual accountability~\cite{le2026}. High verification demands and cognitively demanding explanations can increase susceptibility to
automation bias, while active verification can reduce complacency when evaluation requires high cognitive effort~\cite{romeo2026}. 
A position paper identifies acceptance of vulnerable generated code as an example of overreliance but does not estimate how frequently developers accept insecure code~\cite{ibrahim2025}. Controlled experiments found that ChatGPT assistance improved immediate logical-reasoning scores while participants showed weak metacognitive monitoring and overestimated their AI-assisted
performance~\cite{fernandes2026}.

Temporary offloading does not necessarily produce lasting skill gains. The OECD synthesis reports that several studies found improved task performance during access to generative AI but no corresponding advantage, and sometimes worse performance, when access was removed~\cite{oecdedu2026}. These dynamics resemble the classical ironies of automation, under which removing operators from routine control can weaken monitoring and intervention readiness when they remain responsible for rare failures~\cite{bainbridge1983ironies}. Generative AI expands this vulnerability from physical tasks to abstract symbolic reasoning. An opinion article argues that diagnostic overreliance may threaten the development and retention of clinical pattern-recognition skills~\cite{eltarhouny2026}.

\section{Indirect Evidence of Long-Term Structural Decline} \label{sec:appendix_atrophy_evidence}

This appendix outlines candidate neurobiological and metacognitive mechanisms that have been proposed to underlie human-AI cognitive coevolution, with a focus on possible substrate-level effects and behavioral dependence. We are not aware of longitudinal evidence at the multi-year horizon needed to establish whether any of these mechanisms produce a lasting, let alone irreversible, decline in unassisted human reasoning~\cite{znidaric2026review}. The appendix treats these findings as early evidence and plausible hypotheses.

During an essay-writing task, LLM assistance was associated with lower directed EEG connectivity than search-engine and unaided conditions~\cite{kosmyna2025brainchatgptaccumulationcognitive}.
Among the nine volunteers who switched from LLM-assisted to unaided writing, task-related EEG connectivity in the fourth session differed from the patterns reported during their earlier assisted sessions. The authors interpret this result as short-term cognitive debt.
A subsequent comment, however, notes that the cohorts involved were small and the statistical power limited~\cite{stankovic2025}, so the finding should be treated as preliminary rather than conclusive. If this pattern were to generalize, one interpretation, consistent with Cognitive Load Theory~\cite{sweller1988cognitive} and the ICAP framework~\cite{Chietal2014}, is that offloading reasoning to an external tool reduces the germane cognitive load and engagement ordinarily needed for durable learning; this remains an inference from a single early study rather than a demonstrated mechanism.

Sustained reliance on AI assistance could interfere with the retrieval and consolidation processes that support robust knowledge structures~\cite{Oakley2025}. Oakley draws an analogy with grokking in artificial neural networks, where generalized performance can emerge after extended training beyond memorization. By this analogy, using an external aid before learning is consolidated could leave a person with a less generalizable understanding than sustained unassisted effort would provide~\cite{Oakley2025}. Current evidence does not establish a shared mechanism between biological learning and neural-network grokking or show that AI use causes long-term structural decline.

If sustained reliance reduced knowledge consolidation, it might also widen the gap between objective competence and subjective confidence. The Google effect shows that expecting information to remain externally available shifts memory toward where information can be retrieved rather than the information itself~\cite{sparrow2011}. Messeri and Crockett argue that adopting AI tools in scientific research can create illusions in which researchers believe they understand more than they do~\cite{MesseriCrockett2024}. Whether these effects develop into persistent metacognitive miscalibration remains untested.

Cheng et al. found that 11 leading models affirmed users' actions 49\% more often than human comparators, including when the actions involved deception, illegality, or other harms~\cite{Cheng2026}. A related analysis characterizes social sycophancy across models and open-ended interpersonal contexts~\cite{ChengElephant2025}. These findings motivate the hypothesis that repeated validation could reduce users' practice in tolerating disagreement or exercising social and epistemic accountability. Longitudinal effects on these behaviors have not been established.

\section{Fixed Point Solution} \label{app:stability-fixed-point}
A fixed point $(X^*, Y^*)$ occurs when the latent change scores stop changing over time ($\Delta X_t = 0$ and $\Delta Y_t = 0$). Setting the equations to zero and solving the linear system yields the unique equilibrium state:
\begin{align}
    X^* &= \frac{\gamma_{xy,\text{disc}} \alpha_{y} s_{y} - \beta_{y,\text{disc}} \alpha_{x} s_{x}}{\beta_{x,\text{disc}} \beta_{y,\text{disc}} - \gamma_{xy,\text{disc}} \gamma_{yx,\text{disc}}} \label{fixed-point-disc-x} \\
    Y^* &= \frac{\gamma_{yx,\text{disc}} \alpha_{x} s_{x} - \beta_{x,\text{disc}} \alpha_{y} s_{y}}{\beta_{x,\text{disc}} \beta_{y,\text{disc}} - \gamma_{xy,\text{disc}} \label{fixed-point-disc-y} \gamma_{yx,\text{disc}}}
\end{align}
This solution exists provided that the determinant $\beta_{x,\text{disc}} \beta_{y,\text{disc}} - \gamma_{xy,\text{disc}} \gamma_{yx,\text{disc}} \neq 0$.

To evaluate stability, we rewrite the discrete-time system to express the state at time $t$ using $\Delta X_t = X_t - X_{t-1}$ and $\Delta Y_t = Y_t - Y_{t-1}$:
\begin{align}
    X_t &= \alpha_x s_x + (1 + \beta_{x,\text{disc}}) X_{t-1} + \gamma_{xy,\text{disc}} Y_{t-1} \\
    Y_t &= \alpha_y s_y + \gamma_{yx,\text{disc}} X_{t-1} + (1 + \beta_{y,\text{disc}}) Y_{t-1}
\end{align}

The structural Jacobian matrix $\mathbf{J}$ governing this system is:
\begin{equation}
    \mathbf{J} = \begin{bmatrix} 1 + \beta_{x,\text{disc}} & \gamma_{xy,\text{disc}} \\ \gamma_{yx,\text{disc}} & 1 + \beta_{y,\text{disc}} \end{bmatrix}
\end{equation}
Its trace ($\text{Tr}$) and determinant ($\det$) are:
\begin{align}
    \text{Tr}(\mathbf{J}) &= 2 + \beta_{x,\text{disc}} + \beta_{y,\text{disc}} \\
    \det(\mathbf{J}) &= (1 + \beta_{x,\text{disc}})(1 + \beta_{y,\text{disc}}) - \gamma_{xy,\text{disc}}\gamma_{yx,\text{disc}}
\end{align}

According to the Jury stability criterion for $2 \times 2$ systems, the fixed point is stable (eigenvalues $|\lambda| < 1$) if and only if the following three conditions are simultaneously satisfied. 
The first Jury condition requires the determinant magnitude to be less than 1:
\begin{equation}
    \left| (1 + \beta_{x,\text{disc}})(1 + \beta_{y,\text{disc}}) - \gamma_{xy,\text{disc}}\gamma_{yx,\text{disc}} \right| < 1
\end{equation}
The second Jury condition is:
\begin{equation}
    4 + 2\beta_{x,\text{disc}} + 2\beta_{y,\text{disc}} + \beta_{x,\text{disc}}\beta_{y,\text{disc}} - \gamma_{xy,\text{disc}}\gamma_{yx,\text{disc}} > 0
\end{equation}
The third Jury condition is:
\begin{equation}
    \beta_{x,\text{disc}}\beta_{y,\text{disc}} - \gamma_{xy,\text{disc}}\gamma_{yx,\text{disc}} > 0
\end{equation}

The derived compound inequality is 
\begin{equation}
    0 < \beta_{x,\text{disc}}\beta_{y,\text{disc}} - \gamma_{xy,\text{disc}}\gamma_{yx,\text{disc}} < -\beta_{x,\text{disc}} - \beta_{y,\text{disc}}
\end{equation}

The exact relaxation time for the discrete system is:
\begin{equation}
\tau =-\frac{1}{\ln |\lambda _{\max }|} 
\end{equation}
with
\begin{equation}
\begin{split}
    \lambda_{\max} = \frac{1}{2} \Big[ & 2 + \beta_{x,\text{disc}} + \beta_{y,\text{disc}} \\
    & + \sqrt{(\beta_{x,\text{disc}} - \beta_{y,\text{disc}})^{2} + 4\gamma_{xy,\text{disc}}\gamma_{yx,\text{disc}}} \,\Big]
\end{split}
\end{equation}

\section{Feasible ranges} \label{app:feasible-ranges}

The stability condition for the fixed point is
\begin{equation}
0 < \beta_{x,\text{disc}}\beta_{y,\text{disc}} - \gamma_{xy,\text{disc}}\gamma_{yx,\text{disc}}
< -\beta_{x,\text{disc}} - \beta_{y,\text{disc}} .
\label{eq:stability-1}
\end{equation}

The exact relaxation time is
\begin{equation}
\begin{split}
\tau &= -\frac{1}{\ln|\lambda_{\max}|}, \\
\lambda_{\max} &= \frac{1}{2} \Big[ 2 + \beta_{x,\text{disc}} + \beta_{y,\text{disc}} \\
&\quad + \sqrt{(\beta_{x,\text{disc}} - \beta_{y,\text{disc}})^2 + 4\gamma_{xy,\text{disc}}\gamma_{yx,\text{disc}}} \, \Big].
\end{split}
\label{eq:tau}
\end{equation}

The calibration uses the following domains:
\begin{alignat}{2}
\beta_{x,\text{disc}}   &\in (-0.3,\,0),   &\quad \beta_{y,\text{disc}}   &\in (-0.15,\,0), \\
\gamma_{xy,\text{disc}} &\in (0,\,0.15),   &\quad \gamma_{yx,\text{disc}} &\in (0,\,0.3),   \\
\beta_{x,\text{disc}}   &< \beta_{y,\text{disc}}, &\quad \gamma_{xy,\text{disc}}  &< \gamma_{yx,\text{disc}}.
\end{alignat}

The goal is to find the subset of these domains for which \eqref{eq:stability-1} holds and $\tau_1<\tau<\tau_2$.

Write $a=-\beta_{x,\text{disc}}>0$ and $b=-\beta_{y,\text{disc}}>0$.
Then $\beta_{x,\text{disc}}\beta_{y,\text{disc}}=ab$ and
$-\beta_{x,\text{disc}}-\beta_{y,\text{disc}}=a+b$.
With $P\equiv\gamma_{xy,\text{disc}}\gamma_{yx,\text{disc}}$, condition \eqref{eq:stability-1} becomes
\begin{equation}
ab - a - b < P < ab .
\label{eq:stabP}
\end{equation}

For the lower bound in \eqref{eq:stabP} to be active we would need $ab-a-b>0$,
i.e.\ $(a-1)(b-1)>1$.
But $a<0.3$ and $b<0.15$, so
\begin{equation}
(a-1)(b-1)\in(0.7\times0.85,\,1)=(0.595,\,1),
\end{equation}
and $(a-1)(b-1)>1$ is never satisfied in the prescribed domain.
Since $P>0$ while $ab-a-b=\beta_{x,\text{disc}}\beta_{y,\text{disc}}+\beta_{x,\text{disc}}+\beta_{y,\text{disc}}<0$,
the left inequality in \eqref{eq:stabP} is automatically true.

\begin{center}
\fbox{\parbox{0.95\columnwidth}{
\centering
\textbf{Stability condition:} In the given parameter ranges, stability reduces to
\begin{equation}
\boxed{P \equiv \gamma_{xy,\text{disc}}\,\gamma_{yx,\text{disc}} \;<\; \beta_{x,\text{disc}}\,\beta_{y,\text{disc}} } .
\end{equation}
}}
\end{center}

Introduce the shorthand
\begin{equation}
S = \beta_{x,\text{disc}}+\beta_{y,\text{disc}}, \qquad
\Delta = \beta_{x,\text{disc}}-\beta_{y,\text{disc}} .
\end{equation}
Because $\beta_{x,\text{disc}}<\beta_{y,\text{disc}}$, we have $\Delta<0$.
Equation \eqref{eq:tau} reads
\begin{equation} \label{eq:lambda-max}
\lambda_{\max}(P) = \frac{2+S+\sqrt{\Delta^{2}+4P}}{2}.
\end{equation}
For fixed $(\beta_{x,\text{disc}},\beta_{y,\text{disc}})$, $\lambda_{\max}$ is a
strictly increasing function of $P$.
Its endpoints are
\begin{align}
\lambda_{\max}(P=0) &= 1+\beta_{y,\text{disc}},
\label{eq:lamP0}\\
\lim_{P\to\beta_{x,\text{disc}}\beta_{y,\text{disc}}}\lambda_{\max}(P) &= 1
\quad\Longrightarrow\quad \tau\to\infty .
\end{align}

Let $ \lambda_{\tau_i} = e^{-1/\tau_i}$, $i=1,2$. It follows
\begin{equation} \label{eq:tau2}
\lambda_{\tau_1} < \lambda_{\max} < \lambda_{\tau_2},
\end{equation}
Combining (\ref{eq:lamP0})-(\ref{eq:tau2}), we have 
\begin{equation}
\beta_{y,\text{disc}} \;<\; e^{-1/\tau_2}-1.
\end{equation}
Thus, for example, if $\tau_2=60$, $\beta_{y,\text{disc}} < e^{-1/60}-1 \approx  -0.01653$ and feasible $\beta$ ranges could be calculated as: 
\begin{center}
\fbox{\parbox{0.95\columnwidth}{
\centering
\textbf{Feasible $\beta$ ranges:}
\begin{equation}
\begin{split}
&\boxed{\beta_{x,\text{disc}}\in(-0.30,\,-0.01653)}, \\[4pt]
&\boxed{\beta_{y,\text{disc}}\in(-0.15,\,-0.01653)},
\end{split}
\end{equation}
with $\beta_{x,\text{disc}}<\beta_{y,\text{disc}}$.
}}
\end{center}

From (\ref{eq:lambda-max}), we have 
\begin{equation}
    P = 
\frac{\bigl(2\lambda_{\text{max}}-2-S\bigr)^{2}-\Delta^{2}}{4},\\[4pt]
\end{equation}
Let
\begin{align}
P_{\text{low}} &= \max  \; \Bigl(0,\; f(\tau_1) \Bigr),\\[4pt]
P_{\text{high}} &= \min \; \Bigl(\beta_{x,\text{disc}}\beta_{y,\text{disc}},\; f(\tau_2)
\Bigr),
\end{align}
where 
\begin{equation}
    f(\tau_i) = \frac{\bigl(2e^{-1/\tau_i}-2-S\bigr)^{2}-\Delta^{2}}{4}
\end{equation}
for $i=1,2$. 

For any valid pair $(\beta_{x,\text{disc}},\beta_{y,\text{disc}})$, the product
$P=\gamma_{xy,\text{disc}}\gamma_{yx,\text{disc}}$ must satisfy
\begin{equation}
P_{\text{low}}(\beta_{x},\beta_{y}) \;<\; P \;<\; P_{\text{high}}(\beta_{x},\beta_{y}),
\end{equation}
On the other hand, given a valid $P$, the individual couplings must obey
\begin{equation}
\begin{aligned}
  \gamma_{xy,\text{disc}} \gamma_{yx,\text{disc}} &= P, \\
  \gamma_{xy,\text{disc}} &< \gamma_{yx,\text{disc}}, \\
  \gamma_{xy,\text{disc}} &\in (0, 0.15), \\
  \gamma_{yx,\text{disc}} &\in (0, 0.3).
\end{aligned}
\end{equation}
These imply
\begin{equation}
\boxed{\frac{P}{0.3} \;<\; \gamma_{xy,\text{disc}} \;<\;
\min \; \bigl(0.15,\;\sqrt{P}\,\bigr)},
\end{equation}
and, once $\gamma_{xy,\text{disc}}$ is chosen,
\begin{equation}
\gamma_{yx,\text{disc}} = \frac{P}{\gamma_{xy,\text{disc}}}.
\end{equation}

Across the whole feasible domain, assuming $\tau_1=20$ and $\tau_2=60$, the parameters could be chosen as follows:       
\begin{enumerate}
\item {Choose $\beta$'s.} Pick
$\beta_{y,\text{disc}}\in(-0.15,-0.01653)$ and
$\beta_{x,\text{disc}}\in(-0.30,\beta_{y,\text{disc}})$.

\item {Compute $S$ and $\Delta$.}
$S=\beta_{x,\text{disc}}+\beta_{y,\text{disc}}$,
$\Delta=\beta_{x,\text{disc}}-\beta_{y,\text{disc}}$.

\item {Compute the $P$-window.}
\begin{align*}
P_{\text{low}} &= \max \; \Bigl(0,\;
f(\tau_1) \Bigr),\\
P_{\text{high}} &= \min \; \Bigl(\beta_{x,\text{disc}}\beta_{y,\text{disc}},\; f(\tau_2)
\Bigr).
\end{align*}

\item {Choose $P$.} Select any
$P\in(P_{\text{low}},P_{\text{high}})$.

\item {Choose $\gamma_{xy,\text{disc}}$.} Pick
\begin{equation*}
\gamma_{xy,\text{disc}}\in\Bigl(\frac{P}{0.3},\;
\min(0.15,\sqrt{P})\Bigr).
\end{equation*}

\item Set $\gamma_{yx,\text{disc}}$: 
\begin{equation*}
    \gamma_{yx,\text{disc}}=\frac{P}{\gamma_{xy,\text{disc}}}
\end{equation*}

\end{enumerate}

\section{Composite score} \label{app:composite-score}

We define a two-coordinate normalization from Fluid Intelligence ($X$) and Crystallized Intelligence ($Y$). It is dimensionless before rescaling and serves only as an internal normalization. Validated Woodcock-Johnson IV composites, complete CHC scores, and clinical estimates of general intelligence require separate measurement models. This appendix places the $X$--$Y$ coordinates on a familiar mean-100, standard-deviation-15 presentation scale.

To define this normalization, we combine the standardized $Z-$scores of the components and scale their sum using an assumed correlation:
\begin{equation}\label{eq:composite_z}
Z_{\text{Composite}} = \frac{Z_{X} + Z_{Y}}{\sqrt{2 + 2r}}
\end{equation}
Here, $Z_{\text{Composite}}$ is the standardized constructed score, $Z_{X}$ and $Z_{Y}$ are the standardized component scores, and $r$ is their assumed correlation. The numerical examples set $r=0.65$. Under the assumed standardized two-variable model, the equal-weight sum has standard deviation $\sqrt{2+2r}$. Dividing the sum by this quantity yields a composite with unit variance.

We transform $Z_{\text{Composite}}$ to a conventional presentation scale ($\text{Mean} = 100, \text{SD} = 15$):
\begin{equation}\label{eq:standard_iq}
\text{Composite IQ} = (Z_{\text{Composite}} \times 15) + 100
\end{equation}

For illustration, let the fluid intelligence score ($X$) be 1.0 standard deviation above the mean, the crystallized intelligence score ($Y$) be 2.0 standard deviations above the mean, and $r_{X,Y}=0.65$. With $Z_{X} = 1.0$, $Z_{Y} = 2.0$, and $r_{X,Y} = 0.65$, Eq.~\ref{eq:composite_z} gives:
\begin{align*}
Z_{\text{Composite}} &= \frac{1.0 + 2.0}{\sqrt{2 + 2(0.65)}}  \approx 1.6514
\end{align*}
For $Z_{\text{Composite}} \approx 1.6514$, Eq.~\ref{eq:standard_iq} gives:
\begin{align*}
\text{Composite IQ} &\approx 124.77
\end{align*}
 Rounded to the nearest integer for presentation, the Composite IQ is 125. This constructed score ($125$) exceeds the arithmetic mean of the component standard scores ($122.5$) because the equal-weight sum is normalized by $\sqrt{2+2r}$ before rescaling. The difference is purely algebraic and carries no additional evidential weight concerning trait variance or measurement error.

Example. At Age 60, given a fluid intelligence score of $X = 88$ and a crystallized intelligence score of $Y = 105$, evaluated alongside the assumed correlation coefficient of $r = 0.65$, the constructed cross-lifecycle composite score is exactly 96.15 (which rounds to 96).

The component scores are mapped to standard-deviation units ($SD = 15$) relative to the young-adult peak reference norm ($\text{Mean} = 100$):
\begin{align*}
\text{Fluid Z-Score } (Z_{X}) &= \frac{88 - 100}{15} = -0.8000 \\[1ex]
\text{Crystallized Z-Score } (Z_{Y}) &= \frac{105 - 100}{15} = +0.3333
\end{align*}
The standardized component scores are combined using the population correlation parameter ($r_{X,Y} = 0.65$):
\begin{align*}
Z_{\text{Composite}} &= \frac{Z_{X} + Z_{Y}}{\sqrt{2 + 2r}} \approx -0.2569
\end{align*}
The resulting composite Z-score is transformed to the presentation scale ($\text{Mean} = 100, SD = 15$):
\begin{align*}
\text{Composite IQ} &= (Z_{\text{Composite}} \times 15) + 100 \\
 &=  96.15
\end{align*}

Equivalently, the Composite IQ metric $C$ is the following algebraic function of the component standard scores $X$ and $Y$:
\begin{equation*}
    C = 100 + 15 \cdot \left( \frac{ \frac{X-100}{15} + \frac{Y-100}{15}} {\sqrt{2 + 2r}} \right) = 100 + \frac{X+Y -200}{\sqrt{2+2r}}
\end{equation*}

\section{Theoretical Mapping of Machine AGI Domains to Human Cognitive Frameworks}
\label{app:cognitive_mapping}

This appendix first relates the 10 digital artificial general intelligence (AGI) domains to corresponding CHC broad abilities~\cite{hendrycks2025definitionagi} and then reduces them to two cognitive dimensions. McGrew~\cite{mcgrew2009chc} integrates Carroll's three-stratum model with the Cattell-Horn Gf-Gc theory: Stratum III contains a general intelligence ($g$) factor, Stratum II contains broad ability domains, and Stratum I contains over 80 narrow abilities. The resulting taxonomy comprises 10 primary domains and 6 sensory, motor, and domain-specific domains frequently represented in applied intelligence tests such as the Woodcock-Johnson.

Table 1 in McGrew~\cite{mcgrew2009chc} lists 16 Stratum II broad cognitive abilities across the Cattell-Horn and Carroll models: Fluid Reasoning (Gf), Comprehension-Knowledge (Gc), Domain-Specific Knowledge (Gkn), Visual Processing (Gv), Auditory Processing (Ga), Olfactory Abilities (Go), Tactile Abilities (Gh), Kinesthetic Abilities (Gk), Short-Term Memory (Gsm), Long-Term Storage and Retrieval (Glr), Quantitative Knowledge (Gq), Reading and Writing Ability (Grw), Cognitive Processing Speed (Gs), Decision Speed/Reaction Time (Gt), Psychomotor Speed (Gps), and Psychomotor Abilities (Gp).

 For digital, non-physical intelligence, Hendrycks et al.~\cite{hendrycks2025definitionagi} exclude McGrew's six physical or sensory domains and define ten machine domains from the remaining traits. This adapted digital taxonomy includes General Knowledge (K) for broad factual understanding across diverse subjects, Reading and Writing (RW) for language literacy and comprehension, and Mathematical Ability (M) for quantitative skills spanning arithmetic to calculus. It evaluates problem-solving via On-the-Spot (Fluid) Reasoning (R) alongside active data manipulation via multimodal Working Memory (WM). Memory is further split into Long-Term Memory Storage (MS) for continuous learning and synthesis, and Long-Term Memory Retrieval (MR) for fluently accessing data while minimizing hallucinations. Finally, the framework measures perception and execution through Visual Processing (V) and Auditory Processing (A), while tracking efficiency via Processing Speed (S).

We assign each of the 10 digital domains to the crystallized component ($G_c$), the fluid component ($G_f$), or explicit exclusion. R-to-$G_f$ and K-to-$G_c$ are direct correspondences; the remaining assignments aggregate distinct CHC broad abilities into the two components.

{Domains assigned to the crystallized component ($G_c$)} -- The assignment below is a model reduction; RW, M, MS, and MR correspond to CHC broad abilities distinguishable from $G_c$:
\begin{itemize}
    \item General Knowledge (K): Direct equivalent to crystallized broad comprehension-knowledge, indexing the ultimate volume of verbal and cultural concepts (narrow abilities $K0$ and $K2$).
    \item Reading and Writing (RW): Corresponds to the distinct CHC broad ability $G_{rw}$ and is assigned here to the crystallized component.
    \item Mathematical Ability (M): Corresponds to the distinct CHC broad ability $G_q$ and is assigned here to the crystallized component.
    
    \item Long-Term Memory Storage (MS): Is grouped with the older combined CHC factor $G_{lr}$ and assigned here to the crystallized component.
    \item Long-Term Memory Retrieval (MR): Is grouped with the older combined CHC factor $G_{lr}$ and assigned here to the crystallized component.
    
\end{itemize}

{Domains assigned to the fluid component ($G_f$)} -- R is the clearest direct correspondence; WM and S correspond to distinct CHC broad abilities and are grouped here with the fluid component:
\begin{itemize}
    \item On-the-Spot Reasoning (R): We assign this domain to the fluid component ($G_f$) because it emphasizes inductive and deductive reasoning in novel tasks. This is a modeling correspondence rather than an assertion that R is identical to human fluid intelligence.
    \item Working Memory (WM): Corresponds to the distinct CHC broad ability $G_{sm}$ and is assigned here to the fluid component.
    \item Processing Speed (S): Corresponds to the distinct CHC speed abilities $G_s$ and $G_t$ and is assigned here to the fluid component.
\end{itemize}

The two remaining domains, Visual Processing (V) and Auditory Processing (A), are assigned to neither component. In McGrew's Table~1, they correspond to $G_v$ and $G_a$, respectively, which are Stratum II broad abilities in their own right and are not subsumed under either $G_f$ or $G_c$. Because the model contains only two human variables, we omit domains that belong to neither component. The sensitivity analysis below quantifies the effect of this exclusion.

Taken together, these assignments define the two-dimensional reduction used in the model. R maps to $G_f$, whereas K maps to $G_c$. RW, M, WM, S, MS, and MR correspond to distinct CHC broad abilities and are grouped into the two components according to the functional interpretation described above. Among these modeling choices, the placement of MS and MR has the largest effect on the resulting benchmark.

To assess the sensitivity of the benchmark to these assignments, Table~\ref{tab:assignment-sensitivity} reports $Z_{\mathrm{ref}}$, $R_{\mathrm{ref}}$, and $H$ for the adopted assignment and four alternative variants. These values are computed from the frontier-model scores of Hendrycks et al.~\cite{hendrycks2025definitionagi} using the composite defined in Eq.~(\ref{eq:ratio}). Assigning V and A to either component changes $R_{\mathrm{ref}}$ by at most $0.03$ in either direction, indicating that their exclusion does not drive the benchmark result. By contrast, the treatment of $G_{lr}$ materially affects the benchmark: excluding MS and MR increases $R_{\mathrm{ref}}$ from $0.543$ to $0.719$. These variants represent alternative direct benchmark constructions; the adopted row remains the calibration target reproduced by the simulations.

\begin{table}[tb]
\centering
\caption{{Sensitivity of the benchmark ratio to the domain assignment.
Scores are those of the frontier model tabulated by Hendrycks et
al.~\cite{hendrycks2025definitionagi}; each group is expressed as a percentage of its own maximum and passed through Eq.~(\ref{eq:ratio}) against the human baseline $C^{*}=95.70$. The adopted row is also the calibration target for Eq.~(\ref{eq:Z-max}); the other rows are sensitivity variants, not simulation settings.}}
\label{tab:assignment-sensitivity}
\begin{tabular}{@{}lccccc@{}}
\hline
assignment & $X_{AI}$ & $Y_{AI}$ & {$Z_{\mathrm{ref}}$} & {$R_{\mathrm{ref}}$} & $H$ \\
\hline
adopted, Eqs.~(\ref{eq:assign-gf})--(\ref{eq:assign-gc})
                    & 46.7 & 66.0 & 51.92 & 0.543 & 0.080 \\
V, A into $G_f$     & 48.0 & 66.0 & 52.66 & 0.550 & 0.084 \\
V, A into $G_c$     & 46.7 & 61.4 & 49.41 & 0.516 & 0.066 \\
MS, MR into $G_f$   & 36.0 & 96.7 & 62.93 & 0.658 & 0.158 \\
MS, MR discarded    & 46.7 & 96.7 & 68.81 & 0.719 & 0.211 \\
\hline
\end{tabular}
\end{table}

\section{Steady-state equation} \label{app:fixed-point}

 Setting all time derivatives in Eqs.~(\ref{eq:slow-X})--(\ref{eq:fast-Z}) to zero defines the nonzero equilibria $(\hat{X}, \hat{Y}, \hat{Z})$ through a coupled nonlinear system. The third equilibrium condition is:
\begin{equation*}
\varepsilon\frac{dZ}{dt} = Z \left(1 - \frac{Z}{Z_{\text{max}}}\right) = 0
\end{equation*}
On the nonzero branch ($\hat{Z} \neq 0$):
\begin{equation*}
\hat{Z} = K_0 + \kappa_{x} \hat{X}  + \kappa_{y} \hat{Y}
\end{equation*}
Substituting $\hat{Z}$ into the definition of $R$ gives the equilibrium ratio:
\begin{equation*}
\hat{R} = \frac{\hat{Z}}{\hat{C}} = \frac{K_0 + \kappa_{x} \hat{X}  + \kappa_{y} \hat{Y}}{100 + \frac{\hat{X} + \hat{Y} - 200}{\sqrt{2 + 2r}}}
\end{equation*}

Because $\hat{R}$ depends nonlinearly on both $\hat{X}$ and $\hat{Y}$, the effective interaction coefficients $\hat{\beta}_{i}$ remain functions of the steady-state coordinates:
\begin{equation*}
\hat{\beta}_{i}(\hat{X}, \hat{Y}) = \beta_{i} + \omega_{i} \frac{\hat{R}^4}{\hat{R}^4 + 1} \quad \text{for} \quad i\in \{x,y,xy,yx\}
\end{equation*}
The remaining equilibrium conditions are:
\begin{align*}
\hat{\beta}_{x}(\hat{X}, \hat{Y})\hat{X} + \hat{\beta}_{xy}(\hat{X}, \hat{Y}) \hat{Y} &= -\mu_x  \\
\hat{\beta}_{yx}(\hat{X}, \hat{Y}) \hat{X} + \hat{\beta}_{y}(\hat{X}, \hat{Y}) \hat{Y} &= -\mu_y 
\end{align*}

Define the coefficient determinant:
\begin{equation*}
D(\hat{X}, \hat{Y}) = \hat{\beta}_x(\hat{X}, \hat{Y})\hat{\beta}_y(\hat{X}, \hat{Y}) - \hat{\beta}_{xy}(\hat{X}, \hat{Y})\hat{\beta}_{yx}(\hat{X}, \hat{Y})
\end{equation*}
Cramer's rule gives the following implicit equilibrium system:
\begin{align}
\hat{X} &= \frac{\mu_y  \hat{\beta}_{xy}(\hat{X}, \hat{Y}) - \mu_x  \hat{\beta}_y(\hat{X}, \hat{Y})}{D(\hat{X}, \hat{Y})} \label{eq:x-hat}\\[2ex]
\hat{Y} &= \frac{\mu_x  \hat{\beta}_{yx}(\hat{X}, \hat{Y}) - \mu_y  \hat{\beta}_x(\hat{X}, \hat{Y})}{D(\hat{X}, \hat{Y})} \label{eq:y-hat} \\[2ex]
\hat{Z} &= K_0 + \kappa_{x} \hat{X}  + \kappa_{y} \hat{Y} \label{eq:z-hat}
\end{align}
Because $\hat{X}$ and $\hat{Y}$ enter their own coefficients through $\hat{R}$, the system is implicit and is solved numerically.

The fixed point of Eqs.~(\ref{eq:cont-x})--(\ref{eq:cont-y}) is:
\begin{align}
{X^*} &= \frac{\mu_y  {\beta}_{xy} - \mu_x  {\beta}_y}{{\beta}_x {\beta}_y - {\beta}_{xy} {\beta}_{yx}} \label{eq:x-star}\\[2ex]
{Y^*} &= \frac{\mu_x  {\beta}_{yx} - \mu_y {\beta}_x}{{\beta}_x {\beta}_y - {\beta}_{xy} {\beta}_{yx}} \label{eq:y-star}
\end{align}
Under the transformation in Sec.~\ref{subsect-cont}, the discrete fixed point in Eqs.~(\ref{fixed-point-disc-x})--(\ref{fixed-point-disc-y}) maps to the continuous fixed point in Eqs.~(\ref{eq:x-star})--(\ref{eq:y-star}).

\section{Slow--fast reduction and the fold condition}
\label{app:tikhonov}

This appendix justifies the adiabatic elimination of the fast variable used in Sec.~\ref{sec:results}, records the reduced planar system and its Jacobian, and states the condition by which the folds of Table~\ref{tab:bifurcation} are located.

Rescaling to the fast time $\sigma = t/\varepsilon$ turns
Eq.~(\ref{eq:fast-Z}) into the layer problem
\begin{equation}
\frac{dZ}{d\sigma} = g(Z; X, Y) \equiv Z\left(1 - \frac{Z}{Z_{\max}(X,Y)}\right) ,
\label{eq:layer}
\end{equation}
with $X$ and $Y$ frozen as parameters and $Z_{\max}(X,Y) = K_0 + \kappa_x X + \kappa_y Y$. Its equilibria form the critical manifold, which has the two branches $Z = 0$ and $Z = Z_{\max}(X,Y)$.
Differentiating,
\begin{equation}
\frac{\partial g}{\partial Z} = 1 - \frac{2Z}{Z_{\max}} ,
\label{eq:transversal}
\end{equation}
so on the upper branch $\partial g/\partial Z = -1$ and on the lower branch $\partial g/\partial Z = +1$, in both cases independently of $X$ and $Y$. In the original time these are the transversal eigenvalues $-1/\varepsilon$ and $+1/\varepsilon$. The upper branch is therefore normally hyperbolic and uniformly attracting, and the lower branch is repelling, for every $(X,Y)$ at which $Z_{\max} > 0$.

The condition $Z_{\max}>0$ is satisfied throughout the region of interest. Along the continuation from $K_0 = 0$ to the fold at
$\omega_{yx} = {-1.5}\,\gamma_{yx}$, the smallest value attained is
$Z_{\max} = 33.37$. Tikhonov's theorem then applies for any initial condition with $Z(0) > 0$. The fast variable reaches an
$O(\varepsilon)$ neighbourhood of $Z_{\max}$ in time $O(\varepsilon\ln
\varepsilon^{-1})$, and on the subsequent slow timescale the trajectory of the full system remains $O(\varepsilon)$-close to that of the reduced system obtained by setting $Z = Z_{\max}(X,Y)$.

Substituting the slaved value $Z = Z_{\max}$ into Eq.~(\ref{eq:ratio}) gives the ratio as an explicit function of the slow variables,
\begin{equation}
R(X,Y;K_0) = \frac{K_0 + \kappa_x X + \kappa_y Y}
{100 + \dfrac{X+Y-200}{s}} , \qquad s \equiv \sqrt{2+2r} ,
\label{eq:R-reduced}
\end{equation}
and hence, the four modifiers ${\hat\beta}_i = \beta_i + \omega_i H(R)$ with $H(R) = R^4/(R^4+1)$. Throughout, $\tilde\beta_i$ denotes the increment AI adds to a coefficient, as introduced in Eqs.~(\ref{eq:slow-X})--(\ref{eq:slow-Y}), and $\hat\beta_i$ the resulting total, as in Appendix~\ref{app:fixed-point}. The reduced system is the planar vector field
\begin{align}
\dot X &= F_1(X,Y;K_0) = \mu_x + \hat\beta_x X + \hat\beta_{xy} Y,
\label{eq:red-1} \\
\dot Y &= F_2(X,Y;K_0) = \mu_y + \hat\beta_y Y + \hat\beta_{yx} X.
\label{eq:red-2}
\end{align}
Its equilibria are the $(\hat X, \hat Y)$ of Appendix~\ref{app:fixed-point}, with $\hat Z$ recovered from Eq.~(\ref{eq:z-hat}).

Because $H$ depends on both slow variables, the Jacobian is not simply the matrix of modifiers. From Eq.~(\ref{eq:R-reduced}),
\begin{equation}
\frac{\partial R}{\partial X} = \frac{\kappa_x - R/s}{C} , \qquad
\frac{\partial R}{\partial Y} = \frac{\kappa_y - R/s}{C} ,
\label{eq:dR}
\end{equation}
and $H'(R) = 4R^{3}/(R^{4}+1)^{2}$, so that
$H_X = H'(R)\,\partial R/\partial X$ and
$H_Y = H'(R)\,\partial R/\partial Y$. Collecting the terms in which the modifiers vary,
\begin{equation}
P_1 = \omega_x X + \omega_{xy} Y , \qquad
P_2 = \omega_y Y + \omega_{yx} X ,
\label{eq:P12}
\end{equation}
the Jacobian is
\begin{equation}
\mathbf{J} =
\begin{pmatrix}
{\hat\beta}_x + P_1 H_X & {\hat\beta}_{xy} + P_1 H_Y \\[2pt]
{\hat\beta}_{yx} + P_2 H_X & {\hat\beta}_y + P_2 H_Y
\end{pmatrix} .
\label{eq:jacobian}
\end{equation}
The first term in each entry is the direct effect of the modifier and the second is the feedback through the ratio. Eq.~(\ref{eq:jacobian}) agrees with central differences to better than $10^{-9}$ at every point tested, including at the fold itself.

A saddle-node occurs where an equilibrium exists and the Jacobian is singular. Rather than infer the bifurcation from the failure of a continuation, which cannot distinguish a genuine fold from a loss of numerical tracking, we impose both requirements simultaneously and solve the augmented system
\begin{equation}
F_1(X,Y;K_0) = F_2(X,Y;K_0) = \det \mathbf{J}(X,Y;K_0) = 0
\label{eq:augmented-app}
\end{equation}
for the three unknowns $(X, Y, K_0)$ jointly, using Eq.~(\ref{eq:jacobian}) for the determinant. At
$\omega_{yx} = -1.5\,\gamma_{yx}$ with the remaining $\omega_i$ set to zero, this returns
\begin{equation}
K_0^{*} = 11.9803 , \qquad (X, Y) = (50.9187,\; 32.5864) ,
\label{eq:fold-solution}
\end{equation}
with residual below $7\times 10^{-15}$ and Jacobian eigenvalues
$(-0.327234,\; 4.2\times 10^{-13})$. The second eigenvalue is zero to solver precision while the first is bounded away from zero, so the degeneracy is simple, as a saddle-node requires and as a Hopf point would not permit.

A saddle-node is an interior local maximum of $K_0$ along the branch, and it need not be the global one: past the fold the branch can turn back and climb again toward its end, so a search for the largest $K_0$ the branch attains will report that end rather than the bifurcation. The augmented system above is free of this ambiguity because it imposes $\det \mathbf{J} = 0$ directly.

Two additional checks support the fold classification. First, the equilibrium count changes by two across $K_0^{*}$: at $K_0 = K_0^{*} - 0.5$, the system has a node at $(56.74, 43.98)$ and a saddle at $(45.07, 21.15)$, whereas neither persists above $K_0^{*}$. This change is inconsistent with transcritical or pitchfork bifurcations, which preserve the equilibrium count. Second, the measured scaling $|\mathrm{Re}\,\lambda| \sim \delta^{0.462}$ with $R^2 = 0.999$ in Eq.~(\ref{eq:slowing}) is consistent with the square-root scaling of the saddle-node normal form. Together with the simple zero eigenvalue reported above, these checks support the fold classification.

\section{Numerical verification of the saddle-node bifurcation}
\label{app:caseI-details}

Building on the reduced planar system and fold condition derived above, we numerically verify the saddle-node behavior reported in Sec.~\ref{sec:results}. The analysis considers changes in the two cross-coupling parameters and examines the sensitivity of the resulting bifurcation thresholds.

The fast variable was replaced by $Z_{\max}=K_0+\kappa_xX+\kappa_yY$, leaving the planar system in Eqs.~(\ref{eq:red-1})--(\ref{eq:red-2}). We used $\kappa_x=0.220494$, $\kappa_y=0.311842$, and $R_{\mathrm{ref}}=0.5426$. Folds were located from the augmented condition in Eq.~(\ref{eq:augmented-app}), rather than from the endpoint of numerical continuation. Table~\ref{tab:bifurcation} reports the results when one cross-coupling parameter is varied and the other three $\omega_i$ are held at zero.

\begin{table}[h]
\centering
\caption{Fixed points and saddle-nodes obtained when one offloading parameter is varied, with $R_{\mathrm{ref}}=0.5426$. Changes in $X$ and $Y$ are measured from the unaugmented fixed point $(87.68,104.52)$ at $K_0=0$. A dash indicates that no saddle-node occurs before $Z_{\max}=100$. {Because $X$ and $Y$ decline along the four dashed branches, this state-dependent boundary is reached at $K_0=73.6$--$86.6$, rather than at the fixed-reference value $K_0=48.08$.}}
\label{tab:bifurcation}
{\begin{tabular}{@{}llrrc@{}}
\hline
parameter & strength & $\Delta X$ & $\Delta Y$ & $K_0^{*}$ \\
\hline
$\omega_{yx}$ & $-0.5\,\gamma_{yx}$ & $-4.4\%$  & $-7.1\%$  & --- \\
$\omega_{yx}$ & $-1.0\,\gamma_{yx}$ & $-8.5\%$  & $-13.9\%$ & --- \\
$\omega_{yx}$ & $-1.5\,\gamma_{yx}$ & $-12.4\%$ & $-20.3\%$ & 11.98 \\
$\omega_{yx}$ & $-2.0\,\gamma_{yx}$ & $-16.1\%$ & $-26.4\%$ & 8.36 \\
\hline
$\omega_{xy}$ & $-1.0\,\gamma_{xy}$ & $-9.9\%$ & $-8.6\%$  & --- \\
$\omega_{xy}$ & $-2.0\,\gamma_{xy}$ & $-18.9\%$ & $-16.4\%$ & --- \\
$\omega_{xy}$ & $-3.0\,\gamma_{xy}$ & $-27.4\%$ & $-23.8\%$ & 7.36 \\
\hline
\end{tabular}}
\end{table}

For $\omega_{yx}=-1.5\,\gamma_{yx}$, the saddle enters the positive-$Y$ domain at $K_0=8.84$. The node and saddle therefore coexist for $8.84\leq K_0\leq11.98$, where they meet at the fold reported in Eq.~(\ref{eq:fold-solution}).

The relative sensitivity of $Y$ follows from the equilibrium
budget. At the unaugmented fixed point, the investment term supplies
$\gamma_{yx}X^{*}=14.08$ of the $|\beta_y|Y^{*}=16.21$ needed to maintain $Y$, or $87\%$; $\mu_y$ supplies the remaining $13\%$. If the share assigned to $\mu_y$ is changed while the same unaugmented fixed point is retained, the investment threshold moves from
$-0.890\,\gamma_{yx}$ at a zero share, to
$-1.234\,\gamma_{yx}$ at the calibrated $13.1\%$ share, and to
$-1.866\,\gamma_{yx}$ at a $30\%$ share. The longitudinal aging data fix the equilibrium but do not independently identify this division.

For the two-parameter map, changing $\omega_{xy}$ from zero to $-2\gamma_{xy}$ moves the investment threshold from $-1.234\,\gamma_{yx}$ to $-1.427\,\gamma_{yx}$. Hence equal changes of $-1.3$ times the original coefficients do not produce a saddle-node, even though the investment change alone does. Once a saddle-node exists, however, the scaffolding change can bring it forward: at $\omega_{yx}=-1.5\,\gamma_{yx}$, adding $\omega_{xy}=-\gamma_{xy}$ moves $K_0^{*}$ from $11.98$ to $9.06$.

The recovery-time fit in Eq.~(\ref{eq:slowing}) was obtained from windows approaching the exact saddle-node. The corresponding relaxation time $1/|\mathrm{Re}\,\lambda|$ rises from $28.0$ yr at $\delta=2$ to $431$ yr at $\delta=0.005$.

A search over the normalized box $[-2,2]^4$, restricted pointwise to $Z_{\max}\leq100$ and to psychometrically interpretable values $40\leq X,Y\leq140$, found no Hopf bifurcation or attracting cycle. The expanded search and the verification of the repeating solution in Sec.~\ref{sec:hopf} are reported in Appendix~\ref{app:hopf-details}.

\section{Numerical verification of the Hopf bifurcation}
\label{app:hopf-details}

To locate Hopf points beyond the normalized box examined in the saddle-node analysis, we extended the search to $[-4,4]^4$. We sampled the equilibrium manifold directly by drawing $(X,Y,K_0,\omega_x,\omega_y)$ and recovering the two cross-couplings from the equilibrium conditions. All candidates in both box searches were required to satisfy $Z_{\max}(X,Y;K_0)\leq100$. Candidate unstable foci were checked by direct integration; most trajectories left the positive quadrant rather than approaching a periodic orbit.

To characterize the surrounding parameter region, we fixed
$\omega_y=-0.603\,|\beta_y|$, $\omega_{yx}=+1.100\,\gamma_{yx}$, and
$K_0=42.1$, and varied $\omega_x$ and $\omega_{xy}$. The two Hopf curves meet near $\omega_x=2.08\,|\beta_x|$ and
$\omega_{xy}=-3.67\,\gamma_{xy}$. For $\omega_{xy}/\gamma_{xy}<-3.67$, the upper boundary reaches approximately $3.19\,|\beta_x|$ at $\omega_{xy}=-5.1\,\gamma_{xy}$, whereas the lower boundary remains close to $2.0\,|\beta_x|$. At these states the Hill function is close to saturation, $H=0.966$, and the total fluid self-feedback $\beta_x+\omega_xH$ changes sign near $\omega_x\simeq|\beta_x|/H\simeq2.0\,|\beta_x|$. Figure~\ref{fig:hopf}
takes the one-parameter cut $\omega_{xy}=-3.814\,\gamma_{xy}$ through this region and varies only $\omega_x$. At the lower Hopf point used for the cycle calculation, the four total coefficients are
$\hat\beta_x=+0.228$, $\hat\beta_y=-0.245$, $\hat\beta_{xy}=-0.335$, and $\hat\beta_{yx}=+0.331$. The corresponding Jacobian is
\begin{equation}
{\mathbf{J} = \begin{pmatrix} +0.246 & -0.318 \\ +0.330 & -0.246 \end{pmatrix} ,}
\label{eq:hopf-jacobian}
\end{equation}
The opposite signs of its off-diagonal entries provide the feedback needed for rotation.

For the nonlinear check, we integrated trajectories at $\omega_{xy}=-3.814\,\gamma_{xy}$ while increasing $\delta=\omega_x-\omega_x^H$ above zero. The measured limit-cycle amplitude obeys
\begin{equation}
A \sim \delta^{\,0.4891}, \qquad R^{2} = 0.9999 ,
\label{eq:hopf-scaling}
\end{equation}
The exponent is close to the Hopf normal-form value of $1/2$, and the cycle amplitude approaches zero at onset. In the integrations below
onset, trajectories either returned to the stable equilibrium or left the positive quadrant. The linearized period at the onset point is $29.8$ yr. At the reference value $\omega_x=2.030\,|\beta_x|$ with $\omega_{xy}=-3.814\,\gamma_{xy}$, trajectories initialized inside and outside the orbit converge to the same attracting cycle, whose period is $31.5$ yr.

\bibliographystyle{unsrt}
\bibliography{references_arxiv}

\end{document}